\documentclass[12pt]{article}
\usepackage{amssymb,amsmath}
\usepackage{bm} 
\usepackage{booktabs} 
\usepackage{array}
\usepackage{latexsym}
\usepackage{graphicx}
\usepackage{color}
\usepackage{slashed}
\usepackage{datetime}
\usepackage[nosort]{cite}
\usepackage{verbatim}
\usepackage{enumerate}
\usepackage{makecell} 
\usepackage{chngpage} 
\allowdisplaybreaks
\usepackage{psfrag}
\usepackage{mathtools}
\usepackage{mciteplus}
\usepackage{dsfont}
\usepackage{multirow} 
\usepackage{arydshln} 

\usepackage[colorlinks=true,      linkcolor=blue,      urlcolor=blue,      
filecolor=blue,      citecolor=blue,       pdfstartview=FitH,     
pdfpagemode=UseNone,      bookmarksopen=true]{hyperref}  
\usepackage[all]{hypcap}     

\newcommand{\comm}[1]{} 

\newcommand{\kett}[1]{{\left|{#1}\right\rangle}}

\def\({\left(}
\def\){\right)}
\def\[{\left[}
\def\]{\right]}

\def\pd{{\partial}}

\def\One{{\hbox{ 1\kern-.8mm l}}}

\def\barray{\begin{array}}
	\def\earray{\end{array}}
\def\be{\begin{equation}}
	\def\ee{\end{equation}}
\def\bea{\begin{eqnarray}}
	\def\eea{\end{eqnarray}}
\def\bal{\begin{align}}
	\def\eal{\end{align}}

\def\bt{{\widetilde{b}}}

\def\Gt{{\widetilde{G}}}

\def\Jt{{\widetilde{J}}}

\def\Lt{{\widetilde{L}}}

\def\phit{{\widetilde{\phi}}}
\def\psit{{\widetilde{\psi}}}

\newcommand{\oO}{\overline{\Omega}}

\newcommand{\hb}{{\overline{h}}}
\newcommand{\jb}{{\overline{j}}}
\newcommand{\hbn}{{\overline{h}^{\mathrm{ NS}}}}
\newcommand{\jbn}{{\overline{j}^{\mathrm{ NS}}}}

\numberwithin{equation}{section} 
\allowdisplaybreaks  

\makeatletter
\g@addto@macro\bfseries{\boldmath}
\makeatother

\definecolor{cardinal}{rgb}{0.6,0,0}
\definecolor{darkgreen}{rgb}{0,0.4,0}
\definecolor{golden}{rgb}{0.92, 0.7, 0}
\definecolor{midnight}{rgb}{0, 0, 0.5}
\definecolor{darkblue}{rgb}{0, 0, 0.7}
\definecolor{purple}{rgb}{0.5, 0, 0.5}

\def\IR{\mathbb{R}}

\def\cD{{\cal D}}
\def\cF{{\cal F}}

\def\cL{{\cal L}}
\def\cM{{\cal M}}
\def\cN{{\cal N}}

\def\cft{\widetilde{{\cal F}}}

\def\cO{{\cal O}}

\def\Jt{\tilde{J}}

\def\cLh{\widehat{\mathcal{L}}}
\def\rt{{\widetilde{r}}}

\def\volt{\widehat {\rm vol}_4}

\newcommand{\Jb}{\bar{J}}
\newcommand{\Lb}{\bar{L}}

\def\hb{{\bar{h}}}

\def\tF{{\tilde F}}
\def\tA{{\tilde A}}

\newcommand{\hbr}{{\overline{h}^{\mathrm{ R}}}}
\newcommand{\jbr}{{\overline{j}^{\mathrm{ R}}}}
\newcommand{\hr}{{{h}^{\mathrm{ R}}}}
\newcommand{\jr}{{{j}^{\mathrm{ R}}}}
\newcommand{\ttt}{\widetilde{\Theta}}
\def\at{\tilde{a}}

\usepackage{epsf}
\usepackage{cite}
\usepackage{fancyhdr}
\usepackage{bm}
\usepackage{enumerate} 
	
	\usepackage{empheq} \empheqset{box=\fbox}
	\usepackage[usenames,dvipsnames]{xcolor}

	\numberwithin{equation}{section}  
	\allowdisplaybreaks

	\usepackage{graphicx}
	\usepackage{tikz}
	\usetikzlibrary{decorations.markings}
	\tikzset{->-/.style={decoration={
				markings,
				mark=at position #1 with {\arrow{stealth}}},postaction={decorate}}}
	\usepackage{adjustbox}
	\usepackage{pgfplots}
	\pgfplotsset{compat=1.11}
	\usepgfplotslibrary{fillbetween}
	\usetikzlibrary{intersections}
	\usetikzlibrary{patterns}
	
	\pgfdeclarelayer{bg}
	\pgfsetlayers{bg,main}
	
	\usetikzlibrary{calc}
	\tikzset{
		samples=100,
	}
	\pgfplotsset{compat=1.11}
	
	\pgfmathsetmacro\T{3.14}
	\pgfmathsetmacro\A{0.2}
	\pgfmathsetmacro\N{4}
	\pgfmathsetmacro\D{\N*\T}
	
\begin{document}
		
		
		\begin{flushright}
			
		\end{flushright}
		
		\vspace{3mm}
		
		\begin{center}

			{\Huge{{\bf Vector Superstrata}}}
			
			\vspace{14mm}
			
			{\large
				\textsc	{Nejc \v{C}eplak}} 
	\vspace{12mm}

	\textit{Universit\'e Paris Saclay, CNRS, CEA,\\
		Institut de Physique Th\'eorique,\\
		91191, Gif-sur-Yvette, France}

\medskip

\vspace{4mm} 
%

{\footnotesize\upshape\ttfamily 
	nejc.ceplak @ ipht.fr}\\
\vspace{13mm}

\textsc{Abstract}

\end{center}

\baselineskip=18pt
\begin{adjustwidth}{10mm}{10mm} 

\vspace{1mm}
\noindent
We present the construction of several  microstate geometries of the supersymmetric D1-D5-P black hole in which, within  six-dimensional supergravity, the momentum charge is carried by a vector field. 
The fully backreacted geometries are smooth and horizonless: They are  asymptotically AdS$_3 \times S^3$ with an AdS$_2$ throat that smoothly caps off.
We propose a holographic dual for these bulk solutions and discuss their extension to asymptotically flat space.
In addition, we present several uplifts of the full six-dimensional supersymmetric ansatz to ten-dimensions.
In particular, we show that there exists a frame in which geometries based on vector field momentum carriers are entirely in the NS-sector of supergravity, making them possible starting points for the exploration of stringy black-hole microstates.

\end{adjustwidth}

\thispagestyle{empty}
\clearpage



\baselineskip=14pt
\parskip=1pt

\tableofcontents

\baselineskip=18pt
\parskip=3pt


\section{Introduction}
\label{sec:Intro}

The lack of microstructure at the horizon scale of black holes in General Relativity lies at the heart of the information paradox \cite{Mathur:2008nj, Almheiri:2012rt}. 
While semi-classical gravity can count the number of black-hole microstates
\cite{Heydeman:2020hhw, Boruch:2022tno, Iliesiu:2022kny, LopesCardoso:2022hvc}, it seems to be unable to resolve individual microstates.
At the other extreme, one can reproduce the Bekenstein-Hawking entropy of certain supersymmetric black holes  in string theory by counting relevant brane configurations at weak string coupling \cite{Strominger:1996sh}. 
But this provides only a statistical interpretation of the black-hole entropy at  strong coupling  where we expect classical black holes to exist \cite{Horowitz:1996nw}.

A possible realisation of black-hole microstates is provided by the fuzzball proposal \cite{Mathur:2005zp, Skenderis:2008qn, Bena:2022rna} which states that all microstates of black holes  should be horizonless.%
\footnote{A justification for the absence of horizons is that they have an associated entropy which is incompatible with the properties of pure states.}
A typical microstate is conjectured to possess non-trivial microstructure in the form of a highly quantum and stringy ``fuzz'' that extends to the scale set by the horizon of the corresponding black hole. 
In this context, horizons arise only  by neglecting crucial degrees of freedom \cite{Bena:2022sge} or through an effective description of an ensemble of horizonless microstates.

The fuzzball proposal can be tested  in the D1-D5 system, which is obtained by wrapping $N_1$ D1-branes along a circle $S_y^1$ and $N_5$ D5-branes along $S_y^1\times \cM$, where $\cM$ is a four-dimensional internal manifold that can be either $T^4$ or $K3$. 
For example, the entropy of the supersymmetric two-charge black hole can be reproduced by counting the moduli space of regular supertubes \cite{ Lunin:2002iz, Palmer:2004gu, Rychkov:2005ji}, which are,   through the AdS/CFT correspondence \cite{Maldacena:1997re}, dual to $1/4$-BPS%
\footnote{The fraction is with respect to the total number of supersymmetries of ten-dimensional type IIB supergravity: $1/n$-BPS denotes that the solutions preserves $32/n$ supersymmetries.}
Ramond-Ramond (R-R) ground states in the CFT  \cite{Lunin:2001fv, Lunin:2001jy, Taylor:2005db, Kanitscheider:2006zf, Kanitscheider:2007wq}.
However, the extremal two-charge black hole has a classically vanishing horizon area \cite{Sen:1995in} so generalisations to black-holes with macroscopic horizons are not immediate.
To increase the horizon one  adds a third charge, momentum $P$ along the $S_y^1$ circle.
Microstates of such D1-D5-P black holes have already been extensively analysed, especially those that can be described in supergravity, usually called  microstate geometries \cite{Bena:2013dka, Bena:2022rna}. 

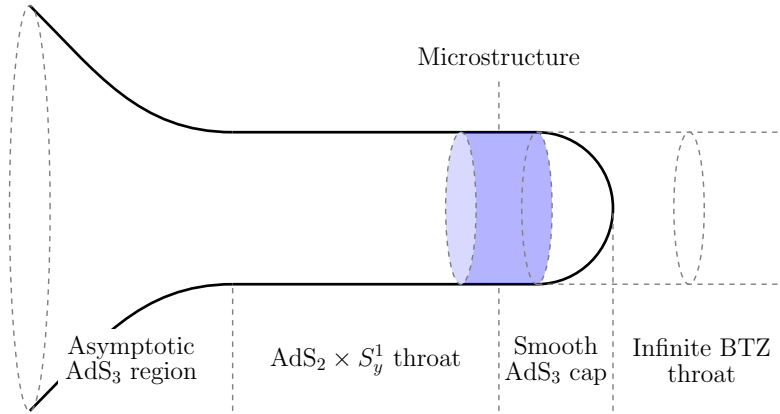
\begin{figure}[t]
	\label{fig:sustra}
	\begin{adjustbox}{max totalsize={0.65\textwidth}{\textheight},center}
		\begin{tikzpicture}
			\begin{scope}[shift={(0,0)}]
				\draw[black,line width=1.5] (0, 1.5) arc (90:-90:1.5);
				\draw [ black, line width=1.5] (0, 1.5) --(-6,1.5);
				\draw [ black, line width=1.5] (0, -1.5) --(-6,-1.5);
				\draw [gray, thick, dashed] (0,1.5) -- (5, 1.5);
				\draw [gray, thick, dashed] (0,-1.5) -- (5, -1.5);
				\draw [gray, thick,dashed, ] (3,0) circle [x radius=0.3, y radius=1.5];
				\filldraw[fill=blue, opacity = 0.3, draw = none] (-1.5,1.5) -- (0,1.5) arc (90:-90: 0.3 and 1.5 ) -- (-1.5,-1.5) arc (-90:90: 0.3 and 1.5 ); 
				\draw [gray, thick,dashed] (1.5, -4) -- (1.5,0);
				\draw [gray, thick,dashed] (-0.75, -4) -- (-0.75,-1.5);
				\draw [gray, thick,dashed] (-0.75, 2.5) -- (-0.75,1.5);
				\draw [gray, thick,dashed] (-6, -4) -- (-6,-1.5);
				\draw [gray, thick,dashed](0.0,0) circle [x radius=0.3, y radius=1.5];
				\filldraw [gray, thick,dashed, fill= blue!15] (-1.5,0) circle [x radius=0.3, y radius=1.5];
				\draw[black, line width=1.5] (-6,-1.5) to[out=180, in=45] (-10, -4);
				\draw[black, line width=1.5] (-6,1.5) to[out=180, in=-45] (-10, 4);
				\draw [gray, thick,dashed](-10,0) circle [x radius=0.4, y radius=4];
				\node[align = center, centered] at (0.375, -3) {\large Smooth\\\large AdS$_3$ cap};
				\node[align = center, centered] at (0.-0.75, 3) {\large Microstructure};
				\node[align = center, centered] at (-3.375, -3) {\large AdS$_2\times S_y^1$ throat};
				\node[align = center, centered, fill=white] at (-8, -3) {\large Asymptotic\\ \large AdS$_3$ region };
				\node[align = center, centered] at (3.25, -3) {\large Infinite BTZ\\ \large throat};
			\end{scope}
		\end{tikzpicture}
	\end{adjustbox}
	\caption{
		A schematic depiction of superstrata.
		In contrast to an extremal black hole, the AdS$_2$ throat has a finite length and is smothly capped off. 
		The details of the microstructure are most prominent at the bottom of the throat (shaded in blue).}
\end{figure}

Superstrata are a particular class of microstate geometries of the D1-D5-P black hole \cite{Bena:2015bea, Bena:2016agb, Bena:2016ypk, Bena:2017xbt, Ceplak:2018pws, Heidmann:2019zws, Heidmann:2019xrd, Shigemori:2020yuo,  Ganchev:2021iwy, Ganchev:2021pgs, Ganchev:2022exf} which have a well-understood CFT interpretation as coherent superpositions of super-descendants of R-R grounds states with non-vanishing momentum charge \cite{ Giusto:2015dfa, Tormo:2019yus, Giusto:2019qig, Rawash:2021pik}.
In some sense, they are $1/8$-BPS generalisations of supertubes. They exhibit all symptoms of a conjectured fuzzball of the D1-D5-P black hole: Approaching from the asymptotic region (either AdS$_3$ or flat space), one  passes through a AdS$_2\times S_y^1$ throat which, unlike in a black hole, is capped off by another AdS$_3$ region, see figure~\ref{fig:sustra}. 
Superstrata are very special representatives of the ensemble of microstates \cite{Raju:2018xue} and give a parametrically subleading contribution to the total entropy of the black hole \cite{Shigemori:2019orj, Mayerson:2020acj}.%
\footnote{See \cite{Wei:2022cpj} for an potential connection between superstrata and disentangled states \cite{Hayden:2020vyo}.}
However, they present a window into possible deviations from the expected behaviour of black holes \cite{Galliani:2017jlg,Bombini:2017sge,Tyukov:2017uig, Bena:2018mpb, Bena:2019azk, Bena:2020iyw, Martinec:2020cml, Ceplak:2021kgl} and can hint at potentially observable features of the underlying microstructure \cite{Mayerson:2020tpn, Bacchini:2021fig, Bah:2021jno, Ikeda:2021uvc}.

The construction of superstrata follows a systematic procedure in which each step on the boundary side has a corresponding action in the bulk. 
Roughly speaking, on the CFT side one starts with a bosonic R-R ground state and then acts on this state with left-moving generators of the symmetry algebra.%
\footnote{By acting only with left-moving generators we preserve some supersymmetry, which allows us to compare states at different points in the moduli space \cite{Baggio:2012rr}.}
In the bulk, this translates to analysing perturbations of the round supertube \cite{Kanitscheider:2006zf, Kanitscheider:2007wq} and acting on them with the bulk representations of the left-moving symmetry generators. 
To determine the backreaction, one needs to solve the BPS equations of the relevant theory of supergravity to all orders in the parameter controlling the amplitude of the perturbation.
Known superstrata are described by six-dimensional $\cN=(1,0)$ supergravity coupled to two tensor multiplets \cite{Gutowski:2003rg,Cariglia:2004kk, Cano:2018wnq}
in which the BPS equations can be organised in a linear and upper-triangular form \cite{Bena:2011dd, Giusto:2013rxa}.
This structure of the BPS equations is crucial in obtaining fully backreacted geometries, which are then dual  to coherent superpositions of bosonic descendants of R-R grounds states with momentum generated by the action of the symmetry algebra.

\subsection{Summary of results}

If the internal manifold is a four-torus,
then one can act with a single supersymmetry generator on a fermionic R-R ground state%
\footnote{Fermionic ground states are absent if the internal manifold is $K3$ in which case there are only 24 bosonic  ground states. Hence throughout this paper we only consider $\cM = T^4$.} \cite{Taylor:2005db}
and obtain another set of bosonic  $1/8$-BPS descendants,  which are associated with breaking the rotation invariance of the $T^4$.
The first result of this paper is the systematic construction of bulk geometries corresponding to coherent superpositions of these $1/8$-BPS states: 
A new class of superstrata in which the momentum charge is carried by vector fields.
Such geometries have an AdS$_2$ throat which is smoothly capped off (exactly as in figure~\ref{fig:sustra}).
With this we initiate the completion of all possible superstrata for the three-charge black hole \cite{Martinec:2022okx}. 

Our bulk construction is guided by the CFT. 
We avoid the use of fermionic excitations of the supertube and the subsequent application of supersymmetry generators by directly analysing  vector excitations around global AdS$_3\times S^3$.
Since such perturbations are in one-to-one correspondence with  $1/8$-BPS  states in the CFT, we use the properties of these states to completely fix the form of the vector excitations.
We show that perturbations obtained in this way satisfy the BPS equations of $\cN=(1,0)$ supergravity coupled to a tensor and a vector multiplet to first order in perturbation theory. 
These BPS equations have recently been shown to have a linear structure  and can be organised into several layers which need to be solved in succession \cite{Ceplak:2022wri}.
This is a critical aspect of our construction, as it  allows us to increase the amplitude of the deformation and construct fully backreacted ``vector'' superstrata.
We will provide a step-by-step procedure on how to obtain such microstate geometries, present several explicit  solutions and analyse their properties.

The second result of this paper is the  uplift of generic BPS configurations of six-dimensional supergravity coupled to a tensor and vector multiplet to solutions of ten-dimensional type II supergravity. 
We provide several different frames and, in particular,  show that there exists an uplift in which only NS fields are excited. 
The pay-off is that one has to isolate a direction of the $T^4$ thus reducing the full ``isotropy'' of the internal manifold to invariance on $T^3$, which is  in stark contrast with solutions based on tensor fields, in which the $T^4$ symmetry is intact.%
\footnote{An exception are superstrata based on internal tensor excitations \cite{Kanitscheider:2007wq,Bakhshaei:2018vux}, but these do not significantly differ from those which are isotropic on $T^4$.}
We thus show that superstrata presented in this paper can be uplifted to purely NS solutions of type II supergravity. As such, they potentially allow for the exact worldsheet analysis and can act as starting points for the construction of more stringy black-hole microstates.

We begin with a brief overview of BPS states in the symmetric product orbifold CFT in Section~\ref{sec:CFT}.
We identify states that are dual to geometries in which, in six-dimensions, the momentum is carried by vector fields. 
In Section~\ref{sec:Perturbation} we analyse vector perturbations of global AdS$_3\times S^3$ and  generate supersymmetric vector excitations of the round supertube.
Section~\ref{sec:Asy-AdS} contains the systematic procedure of how to solve the BPS equations starting with the constructed vector perturbations to obtain fully backreacted, asymptotically AdS$_3\times S^3$ superstrata.
In Section~\ref{sec:ExpExam} we present complete solutions for two families of vector superstrata.
We analyse the properties of generic solutions, including the corresponding asymptotic charges and find agreement with those predicted by the CFT analysis.
In Section~\ref{sec:ato0} we show that in a particular corner of parameter space, some geometries develop an infinitely long AdS$_2$ throat and a horizon.
We discuss why the solutions degenerate and provide possible resolutions \cite{Bena:2022sge}.
In Section~\ref{sec:10D uplift} we analyse how to uplift six-dimensional BPS configurations into ten-dimensional supergravity.
In particular, we show that there exists a frame in which solutions based on vector fields lie entirely in the NS-NS sector, with all R-R gauge fields vanishing.
In Section~\ref{sec:Discussion} we conclude with a few remarks.

In the appendices we collect some technical details. 
In Appendix~\ref{sec:BPSeq} we summarize the ansatz for all supersymmetric solutions of $\cN=(1,0)$ supergravity coupled to a vector and tensor multiplet. We also present the associated BPS equations and how they organise into layers.
In Appendix~\ref{app:10to6} we show, using a particular uplift, that solutions to the equations of motion of six-dimensional supergravity are also classical solutions of ten-dimensional type IIA supergravity. 
Appendix~\ref{app:STdual} contains details of several  ten-dimensional uplifts that are omitted in the main text, while in Appendix~\ref{app:Examples} we present additinal solutions to the BPS equations.
In Appendix~\ref{app:SecondState} we show how to match the conserved charges for an alternative set of states, not considered in the main text.

\section{Analysis of states in the D1-D5 CFT}
\label{sec:CFT}

The starting point of our construction of superstrata will be the analysis of vector perturbations around AdS$_3 \times S^3$.
To determine the exact form of these perturbations, we use the fact that they are in one-to-one correspondence with states in the dual field theory. 
In this section we introduce the D1-D5 CFT at the symmetric product  orbifold point.
We analyse the spectrum NS-NS (anti)-chrial primary states and their descendants, and relate them to states in the R-R sector, using the spectral flow transformation.
In particular, we identify the states which are dual to vector field excitations in six-dimensional supergravity and describe their properties.
We draw heavily from the reviews \cite{Avery:2010qw, Shigemori:2020yuo}. 
A similar analysis was recently done in \cite{Martinec:2022okx}.

\subsection{D1-D5 CFT at the symmetric product orbifold point}

The AdS/CFT correspondence states the Type IIB string theory on  is dual to a two-dimensional supersymmetric CFT, which we call D1-D5 CFT. 
The moduli space of this CFT contains the symmetric product orbifold point where the theory is given in terms of a $N = N_1\,N_5$ copies of free sigma models identified under permutation symmetry, making the total target space $\left(T^4\right)^N/S_N$ and total central  central charge $c = \bar c = 6N$.

The supersymmetry group of the CFT is $SU(1,1|2)_L \times SU(1,1|2)_R$. 
We denote the left-moving affine generators by 
$L_n$, $J^i$, $G_n^{\alpha A}$ and their right-moving analogues by $\Lt_n$, $\Jt^{\bar i}$, and $\Gt_n^{\dot \alpha A}$.  
The $\alpha, \dot \alpha = \pm$ are doublet and $i, \bar i = 1,2,3$ are triplet indices of the R-symmetry group $SU(2)_L\times SU(2)_R\subset SU(1,1|2)_L \times SU(1,1|2)_R$.
The index $A = 1,2$ denotes a doublet of the $SU(2)_B$ outer automorphism symmetry of the superalgebra.
There is an additional custodial group $SU(2)_C$, whose doublet indices we are going to label with $\dot A= 1,2$. 
This group, together with the outer automorphism symmetry combines into the symmetry of the internal four-torus $SU(2)_B \times SU(2)_C \simeq SO(4)_I$. 

We begin by analysing $1/4$-BPS states in the NS-NS sector of the theory.
There are 8 bosonic and 8 fermionic anti-chiral primary states%
\footnote{One can equally work with chiral primary states. We choose to work with anti-chiral primaries to make contact with previous work in the superstrata literature.}, which we denote by \cite{Shigemori:2020yuo}
\begin{subequations}
	\label{eq:ACP}
	\begin{align}
		&\kett{ \alpha \dot \alpha}^{\rm NS}_k\,,  &&h^{\rm NS} = - j^{\rm NS} = \frac{k + \alpha}{2}\,,\qquad  && \hbn = - \jbn =  \frac{k + \dot \alpha}{2}&& \text{boson}\,,\label{eq:nsbos1}\\
		&\kett{ \alpha \dot A}^{\rm NS}_k\,,  &&h^{\rm NS} = - j^{\rm NS} = \frac{k + \alpha}{2}\,, \qquad && \hbn = - \jbn =  \frac{k }{2}\,, && \text{fermion}\,,\label{eq:nsfer1}\\
		&\kett{ \dot A \dot \alpha}^{\rm NS}_k\,, &&h^{\rm NS} = - j^{\rm NS} = \frac{k }{2}\,,\qquad  && \hbn = - \jbn =  \frac{k + \dot \alpha}{2}&& \text{fermion}\,,\label{eq:nsfer2}\\
		&\kett{ \dot A  \dot B }^{\rm NS}_k\,,  &&h^{\rm NS} = - j^{\rm NS} = \frac{k }{2}\,, \qquad && \hbn = - \jbn =  \frac{k }{2}&& \text{boson}\,\label{eq:nsbos2}.
	\end{align}
\end{subequations}
We have also given the eigenvalues $h^{\rm NS}$ and $j^{NS}$ under the action of $L_0$ and  $J_0^3$, and their right-moving analogues.
The integer $k = 1,2,\ldots, N$, denotes the twist sector of the state.
The state with the lowest possible conformal weights, $\kett{--}^{\rm NS}_1$, is the vacuum state and is dual to global AdS$_3\times S^3$. 
All other states in this sector can  be considered as single-particle states and are related through the AdS/CFT correspondence to perturbations around global AdS$_3\times S^3$.
For example, the $T^4$ invariant state
\begin{align}
	\label{eq:00State}
	\kett{00}_k^{\rm NS}\equiv \frac{1}{2}\,\epsilon_{\dot A \dot B}\, \kett{\dot A \dot B}_k^{\rm NS}\,,  &&h^{\rm NS} = - j^{\rm NS} = \frac{k }{2}\,, \qquad && \hbn = - \jbn =  \frac{k }{2}\,,
\end{align}
where $\epsilon_{\dot A \dot B}$ is an anti-symmetric two-tensor, is the perturbation on which most of the known superstrata are built \cite{Bena:2015bea, Bena:2016agb, Bena:2016ypk, Bena:2017xbt, Ceplak:2018pws}.

\subsection{CFT states related to vector fields}

Since anti-chiral primaries are the lowest-weight states with respect to the rigid symmetry group of the CFT, one can create more supersymmetric states through repeated action of $L_{-1}$, $J_0^+$, and $G_{-\frac12}^{+A}$ generators, while leaving the right-moving sector untouched.
We focus only on bosonic descendants, as these can be described by bosonic fields in the bulk.
They can be obtained by starting with a bosonic anti-chiral primary and then acting on that state repeatedly with $L_{-1}$ and $J_0^+$ \cite{Bena:2015bea, Bena:2016agb, Bena:2016ypk, Bena:2017xbt}. 
Alternatively, one can begin by acting  twice with the supersymmetry generator $G_{-\frac12}^{+A}$, which produces a different lowest-weight state on which one can with $L_{-1}$ and $J_0^+$ \cite{Ceplak:2018pws}.

However, one can also make bosonic descendants by starting with a fermionic anti-chiral primary on which one acts with a supersymmetry generator only once. 
Eight such states are obtained by acting with the supersymmetry generator on \eqref{eq:nsfer2}
\begin{subequations}
	\label{eq:states1}
	\begin{align}
		&G_{-\frac12}^{+A}\kett{ \dot A +}_k^{\rm NS} &&h^{\rm NS} = \frac{k}{2}+ \frac12\,, \quad j^{\rm NS} = -\frac{k}{2}+  \frac12\,,&& \hbn = - \jbn =  \frac{k }{2}+  \frac12\,,\label{eq:StateChi+}\\*
		&G_{-\frac12}^{+A}\kett{\dot A -}_k^{\rm NS} &&h^{\rm NS} = \frac{k}{2}+ \frac12\,, \quad j^{\rm NS} = -\frac{k}{2}+ \frac12\,,&& \hbn = - \jbn =  \frac{k }{2}-  \frac12\,,\label{eq:StateChi-}
	\end{align}
\end{subequations}
while the remaining eight are obtained by starting with the other fermionic state \eqref{eq:nsfer1}
\begin{subequations}
	\label{eq:states2}
	\begin{align}
		\label{eq:StartState}
		&G_{-\frac12}^{+A}\kett{+ \dot A}_k^{\rm NS} &&h^{\rm NS} = \frac{k}{2}+1\,, \quad j^{\rm NS} = -\frac{k}{2}\,,&& \hbn = - \jbn =  \frac{k }{2}\,,\\
		&G_{-\frac12}^{+A}\kett{- \dot A}_k^{\rm NS} &&h^{\rm NS} = \frac{k}{2}\,, \quad j^{\rm NS} = -\frac{k}{2}+1\,,&& \hbn = - \jbn =  \frac{k }{2}\,.\label{eq:StatePsi-}
	\end{align}
\end{subequations}
These states are no longer an anti-chiral primaries, however they still annihilated by some of the bosonic symmetry generators
\begin{align}
	\label{eq:StateProperty}
	J_0^-\,\kett{\psi}^{\rm NS}=  \Jb_0^-\,\kett{\psi}^{\rm NS} = L_1\,\kett{\psi}^{\rm NS} = \Lb_1\,\kett{\psi}^{\rm NS}
	= 0\,.
\end{align}
They have an index in the $SU(2)_B \times SU(2)_C \simeq SO(4)_I$ and
are related to vector field excitations  in the six-dimensional picture.
An intuitive way to see this is by recalling that vector fields in six-dimensions arise, for example, by dimensional reduction on $T^4$ when ten-dimensional fields have non-trivial components along a specific direction of the four-torus.\footnote{This can be seen explicitly in Section~\ref{sec:10D uplift}.}
In the CFT, directions along the $T^4$ are characterised by vector indices $a =1,2,3,4$ of $SO(4)_I$, which can be decomposed into $SU(2)_B \times SU(2)_C$ doublets using Pauli matrices $X^{\dot A A} \sim  X^a\, (\sigma^a)^{\dot A A}$, or vice-versa, one can use this basis to convert an object with an $A$ and $\dot A$ index into a $SO(4)_I$ vector \cite{Avery:2010qw}. 
Thus we interpret states with $A$ and $\dot A$ indices to single out a direction in the four-torus and will be, in the six-dimensional picture, described by  vector fields.

Notice that the charges of the states in \eqref{eq:states1} and \eqref{eq:states2} are pairwise related through a shift in the parameter $k$:
Taking the state \eqref{eq:StateChi+} and shifting $k\rightarrow k-1$ yields the charges \eqref{eq:StatePsi-}, while taking \eqref{eq:StateChi-} and sending $k \rightarrow k+1$ gives the charges of \eqref{eq:StartState}.

In this paper, we focus on the state \eqref{eq:StartState} and leave the analysis of the others for future work \cite{wip}.
As can be seen from \eqref{eq:StateProperty}, the state is again a lowest-weight state of the $SU(2)_L$ and $SL(2, \mathbb{R})_L$ multiplets. 
The most general descendant state can be generated by the action of left-moving symmetry generators 
\begin{align}
	\label{eq:NSstate}
	\kett{k,m,n;\dot A, A}^{\rm NS} \equiv \left(L_{-1}\right)^{n-1} \left(J_0^{+}\right)^mG_{-\frac12}^{+A}\kett{+ \dot A}_k^{\rm NS}\,,
\end{align}
with quantum numbers 
\begin{align}
	\label{eq:chargesstate1}
	&&h^{\rm NS} = \frac{k}{2}+n\,, \qquad j^{\rm NS} = -\frac{k}{2}+m\,,&& \hbn = - \jbn =  \frac{k }{2}\,, 
\end{align}
where $n = 1, 2, \ldots$ and $m = 0, 1, \ldots, k$.

The full CFT state consist of a large number of copies of CFT states, possibly in different twist sectors.
Consider a state made by  combining many copies of the vacuum and \eqref{eq:NSstate} 
\begin{align}
	\label{eq:FullStateNS}
	\left(\kett{--}_1^{\rm NS}\right)^{N_a}\, \left(\kett{k,m,n; \dot A,A}^{\rm NS} \right)^{N_b}\,, 
\end{align}
which is subject to the total winding constraint
\begin{align}
	\label{eq:StrandBudget}
	N = N_a + k \, N_b\,.
\end{align}
If we set $N_b =1$, then the charges of the full state are equal to those of the state \eqref{eq:NSstate}. 
In the bulk sich a state corresponds to an $1/8$-BPS perturbation around global AdS$_3 \times S^3$ with charges given by \eqref{eq:chargesstate1}.
By increasing $N_b$, we increase the amplitude of the perturbation. When $N_b \sim N$, the backreaction on the geometry has to be taken into account and the full equations of motion need to be solved.

Microstate geometries are dual to states in the Ramond-Ramond (R-R) sector of the theory, whose conformal dimensions scale with the central charge $c$. 
One can map anti-chiral primaries and their descendants to states in the R-R sector through the spectral flow transformation, under which the charges transform as
\cite{Avery:2010qw, Shigemori:2020yuo}%
\footnote{To map an anti-chiral primary to a Ramond state, one sets $\eta= \frac12$, while taking $\eta = - \frac12$ corresponds to the inverse mapping.}
\begin{align}
	\label{eq:SpectralFlowCFT}
	h' ~=~ h + 2\,\eta\,j + k \, \eta^2\,, \qquad j'~=~  j + k\, \eta\,.
\end{align}
For example, the NS-NS vacuum maps to 
\begin{align}
	\label{eq:vacuumtrans}
	&\kett{--}_1^{\rm NS} \mapsto \kett{++}_1^{\rm R}\,, && \hr = \hbr = \frac14\,, && \jr = \jbr = \frac{1}{2}\,.
\end{align}
The state composed of only $\kett{++}_1^{\rm R}$ is dual to the maximally spinning round supertube \cite{Kanitscheider:2007wq}.
On the other hand, the spectral flow to the state \eqref{eq:NSstate} is
\begin{align}
	\label{eq:Rstate}
	\kett{k,m,n; \dot A, A}^{\rm R} \equiv \left(L_{-1}-J_{-1}^3\right)^{n-1}\, \left(J_{-1}^+\right)^m \, G_{-1}^{+A}\,\kett{-\dot A}_k^{\rm R}\,,
\end{align}
with quantum numbers 
\begin{align}
	h^{R} = \frac{k}{4} + m + n\,, \qquad j^{\rm R} = m\,, \qquad \hb^{\rm R} = \frac{k}4\,,\qquad  \jb^{\rm R} = 0\,.
\end{align}

The full state  \eqref{eq:FullStateNS} is then mapped to 
\begin{align}
	\label{eq:FullStateR}
	\left(\kett{++}_1^{\rm R}\right)^{N_a}\, \left(\kett{k,m,n; \dot A, A}^{\rm R} \right)^{N_b}\,,
\end{align}
subject to the constraint \eqref{eq:StrandBudget}. The eigenvalues are given by 
\begin{align}
	\label{eq:FullRCharges}
	h^{\rm R} =  \frac{N}4 + N_b \left(m+n\right)\,, \qquad \hb^{\rm R} = \frac{N}4\,, \qquad j^{\rm R} = \frac{N_a}{2} + N_b\, m\,, \qquad \jb^{\rm R} = \frac{N_a}{2}\,,
\end{align}
which, most importantly implies that the state and thus the corresponding geometry has non-vanishing momentum
\begin{align}
	\label{eq:MomChargeCFT}
	n_P^{\rm R} \equiv  h^{\rm R}- \hb^{\rm R} =  N_b \left(m+n\right)\,.
\end{align} 
We see that the conformal dimensions scale as $\hr \sim \cO(N) \sim \cO(c)$.
As already mentioned, when $N_b= 0$,  the dual  geometry is  the maximally spinning round supertube. 
When $N_b$ is  small, but finite, the state \eqref{eq:FullStateR} describes a perturbation on the supertube. Further increasing the value of $N_b$ results in a backreacted geometry -- superstrata.%
\footnote{More precisely,  superstrata  are dual to a coherent sum of terms such as \eqref{eq:FullStateR}  \cite{Skenderis:2006ah}. However, for large $N$ this sum is sharply peaked and thus the corresponding charges will be given, to leading order in large $N$, by \eqref{eq:FullRCharges} \cite{Bena:2017xbt}.}
Up to the details of the coherent sum, the state \eqref{eq:FullStateR} is our proposal for  dual of vector field superstrata constructed in this paper. 
As a first check, in a fully backreacted geometry  we calculate the charges independently by a  supergravity calculation and we find complete agreement with the values  \eqref{eq:FullRCharges} and \eqref{eq:MomChargeCFT}.

\section{Vector perturbations}
\label{sec:Perturbation}

In this section, we present the construction of a linear excitation around global AdS$_3\times S^3$.
We fix the form of the perturbation by demanding that it has the same properties with respect to the symmetry algebra generators as the CFT states discussed in the previous section.
We finally use the spectral flow transformation to create the corresponding vector excitation on the round supertube.

\subsection{Round supertube profile}

One of the basic entries in the holographic dictionary is that the state consisting of only $\kett{++}_1^{\rm R}$ is dual to the maximally spinning round supertube \cite{Lunin:2001fv, Lunin:2001jy, Taylor:2005db, Kanitscheider:2007wq}.
To describe this solution it is convenient to parameterize the four-dimensional base space as 
\begin{align}\label{eq:BaseCoords}
	x_1 + i x_2 = \sqrt{r^2 + a^2}\sin\theta\, e^{i\phi}, \qquad x_3 + i x_4 = r \cos\theta\, e^{i\psi},
\end{align}
where $\theta\in[0,{\pi\over 2}]$ and $\phi,\psi\in[0,2\pi)$. 
In these coordinates, the flat metric on $\mathbb{R}^4$ is given by
\begin{align}
	\label{eq:BaseSpace}
	ds_4^2 = \Sigma\left(\frac{dr^2}{a^2 + r^2} + d\theta^2\right) + \left(a^2+ r^2\right)\, \sin^2\theta\, d\phi^2 + r^2 \, \cos^2\theta \, d\psi^2\,,  
\end{align}	 
where we defined
\begin{align}
	\Sigma \equiv r^2 + a^2 \, \cos^2\theta\,.
\end{align}
The round supertube is then described by the following ansatz quantities%
\footnote{For more details on the notation, see Section~\ref{sec:BPSeq}.}
\begin{subequations}
	\label{eq:Supertube}
	\begin{align}
		&Z_1 = \frac{Q_1}{\Sigma}\,,\qquad 
		Z_2 = \frac{Q_5}{\Sigma}\,,\qquad
		&&\gamma_{1,2} = - Q_{1,5} \frac{(r^2 + a^2)\cos^2\theta}{\Sigma}d\phi \wedge d\psi,\\
		&\beta \equiv \beta_0= \frac{R_y\,a^2}{\sqrt{2}\,\Sigma}\,(\sin^2\theta\, d\phi - \cos^2\theta\,d\psi)\,,
		\qquad 
		&&\omega\equiv \omega_0=\frac{R_y\,a^2}{\sqrt{2}\,\Sigma}\,(\sin^2\theta\, d\phi + \cos^2\theta\,d\psi) \,,
	\end{align}
\end{subequations}
with all other quantities vanishing. 
The location of the supertube is at $\Sigma =0$ and despite the harmonic functions $Z_1$ and $Z_2$ diverging at this locus, the metric stays regular provided that  
\begin{align}
	\sqrt{Q_1\, Q_5} = a \, R_y\,,
\end{align}
where $R_y$ is the radius of the $S_y^1$ circle.

As on the CFT side, the round supertube is related to global AdS$_3\times S^3$ through a spectral flow transformation, which now takes a form of a change of coordinates
\begin{align}
	\label{eq:SpectralFlow}
	\phit = \phi - \frac{t}{R_y}, \qquad \psit = \psi - \frac{y}{R_y}.
\end{align}
Indeed, inserting  \eqref{eq:Supertube} into the  ansatz for the metric and using \eqref{eq:SpectralFlow} gives
\begin{align}
\label{eq:GlobalAdS3S3}
	ds^2 =- \;\! \frac{r^2+a^2}{a^2 R_y^2}\,dt^2 +\frac{r^2}{a^2 R_y^2}\,dy^2 
	+\frac{dr^2}{r^2+a^2}
	+d\theta^2+\sin ^2\theta\, d\phit^2+\cos ^2\theta \,d\psit^2\,,
\end{align}
with the corresponding three form $G$ appropriately factorising into the sum of the two volume forms: This is  global AdS$_3\times S^3$ with equal radii $R_{\rm AdS_3}^2 = R_{\rm S^3}^2 = \sqrt{Q_1\,Q_5}$.

\subsection{Building up a perturbation}

We now want to construct a vector excitation on global AdS$_3\times S^3$ that solves the BPS equations to linear order in the perturbation amplitude, $\bt$. 
Since the metric is in product form and depends only on the coordinates $r$ and $\theta$ we make an ansatz \cite{Deger:1998nm}
\begin{align}
	\label{eq:PertAns}
	A^{\rm NS} = \bt\, \Big[f(r) \, A_{\rm S^3}(\theta) + g(\theta) \, A_{\rm AdS_3}(r)\Big]\, e^{i\left( m_1 \phit + m_2\psit + n_1\,\frac{t}{R_y} + n_2\,\frac{y}{R_y}\right)}\,,
\end{align} 
where $f(r)$ and $g(\theta)$ are functions while $ A_{\rm AdS_3}(r)$ and $ A_{\rm S^3}(\theta)$ are one-forms on AdS$_3$ and $S^3$ respectively. 
We want to build the perturbation corresponding to the CFT state \eqref{eq:FullStateNS} with $m=0$ and $n=1$, which is the lowest-weight state in its multiplet at given $k$ and satisfies \eqref{eq:StateProperty}.
In global AdS$_3 \times S^3$ a representation of the symmetry generators is known. The Virasoro algebra is realised by \cite{Maldacena:1998bw}
\begin{subequations}
	\begin{gather}
		\begin{split}
			L_0&={i R_y \over 2}(\partial_t+\partial_y),\\
			L_{\pm 1}
			&=ie^{\pm {i\over R_y}(t+y)}
			\biggl[
			-{R_y\over 2}\biggl({r\over \sqrt{r^2+a^2}}\partial_t+{\sqrt{r^2+a^2}\over r}\partial_y\biggr)
			\pm {i\over 2}\sqrt{r^2+a^2}\,\partial_r
			\biggr],\\
			\Lb_0 &= \frac{i R_y }{2}\left( \partial_t - \partial_y\right)\,,\\
			\Lb_{\pm1}&=  i e^{\pm {i\over R_y}(t-y) } \biggl[
			-{R_y\over 2}\biggl({r\over \sqrt{r^2+a^2}}\partial_t-{\sqrt{r^2+a^2}\over r}\partial_y\biggr)
			\pm {i\over 2}\sqrt{r^2+a^2}\,\partial_r
			\biggr],
		\end{split} 
	\end{gather}
\end{subequations}
while the R-symmetry generators are given by \cite{Giusto:2013bda}
\begin{subequations}
	\label{eq:SU(2)gen}
	\begin{align}
		&J_0^3=-{i\over 2}(\partial_{\phit}+\partial_{\psit}),&&
		J_0^\pm ={1\over 2}e^{\pm i(\phit+\psit)}
		(\pm  \partial_\theta+i\cot\theta\, \partial_{\phit}-i\tan\theta\, \partial_{\psit})\,,\\
		&\Jb_0^3 = - \frac{i}{2}\left( \partial_\phit - \partial_\psit\right)\,,&& 
		\Jb^{\pm}_0 = \frac{1}{2}e^{\pm i (\phit - \psit)} \left( \mp \partial_\theta - i \cot\theta\, \partial_\phit - i \tan\theta \,\partial_\psit\right)\,.
	\end{align}
\end{subequations}
The perturbation should have the same properties as the CFT state, so we impose the conditions
\begin{subequations}
	\label{eq:PertCons}
	\begin{gather}
		L_0\, A_{\rm NS} = \frac{k+2}{2}\,A_{\rm NS}\,,\qquad    \Lb_0 \, A_{\rm NS} = - J_0^3 \, A_{\rm NS}= - \Jb^3_0 \, A_{\rm NS} = \frac{k}{2}\,A_{\rm NS}\,,\\
		J_0^- \, A_{\rm NS} =  \Jb_0^- \, A_{\rm NS} = L_1\,A_{\rm NS}= \Lb_1 \,A_{\rm NS} = 0\,.
	\end{gather}
\end{subequations}
These are enough to fully determine the perturbation%
\footnote{An additional term is  set to zero by demanding that the spectral flowed perturbation has no $du$ component, so that it is consistent with the supersymmetric form \eqref{eq:Potentials}.}
\begin{align}
	A_{\rm NS} = \bt \, \Delta_{k,0,1}\,e^{-i \left( \frac{t + y}{R_y} + k\left(\phit+\frac{t}{R_y}\right)\right)}\left[ \frac{i\,a^2}{r\left(a^2+ r^2\right)} \, dr +  \frac{dt + dy}{R_y}\right]\,, 
\end{align}
where
\begin{align}
	\Delta_{k,m,n} &\equiv
	\left(\frac{a}{\sqrt{r^2+a^2}}\right)^k
	\left(\frac{r}{\sqrt{r^2+a^2}}\right)^n 
	\cos^{m}\theta \, \sin^{k-m}\theta \,.
\end{align}
By acting on this perturbation with $L_{-1}$ and $J_0^+$ generators we can construct perturbations dual to all states in the multiplet
\begin{align}
	A^{(k,m,n)}_{\rm NS} = \bt \, \Delta_{k,m,n}\,e^{-i \left(n \frac{t + y}{R_y} + k\left(\phit+\frac{t}{R_y}\right)- m (\phit + \psit)\right)}\left[ \frac{i\,a^2}{r\left(a^2+ r^2\right)} \, dr +  \frac{dt + dy}{R_y}\right]\,,
\end{align}
which should correspond to the state \eqref{eq:FullStateNS} with the same mode numbers $(k,m,n)$.

Ultimately, we are interested in the R-R sector, where the perturbation is 
\begin{align}
	A^{(k,m,n)}_{\rm R} = \bt\,e^{- i \hat v_{k,m,n}}\, \Delta_{k,m,n}\, \left( \frac{\sqrt{2}}{R_y}\, dv  + \frac{i\,a^2}{r\left(a^2+ r^2\right)} \, dr\right) \,,
\end{align}
or, using the real part only
\begin{align}
	\label{eq:PertRkmn}
	A^{(k,m,n)}_{\rm R} = \bt\, \Delta_{k,m,n}\, \left( \frac{\sqrt{2}}{R_y}\, dv\,\cos{\hat v_{k,m,n}}+ \frac{a^2}{r\left(a^2+ r^2\right)}\, dr\, \sin{\hat v_{k,m,n}} \right) \,,
\end{align}
where we used the phase
\begin{align}
	\label{eq:vhatPhase}
	\hat{v}_{k,m,n} &\equiv (m+n) \frac{\sqrt{2}\,v}{R_y} + (k-m)\phi - m\psi \,.
\end{align}
As we will see in later sections, this is the field that carries the momentum in the new superstrata. 
Indeed, one can show that \eqref{eq:PertRkmn} solves the BPS equations to linear order in $\bt$.
When $\bt$ is increased, other fields get excited as well causing the geometry to backreact and develop a finitely-sized AdS$_2$ throat.

\section{Fully backreacted geometries}
\label{sec:Asy-AdS}

In this section we show how to systematically solve the BPS equations exactly in $\bt$ for single-mode superstrata, starting with the perturbations constructed in the previous section. 
While one can use either of the parametrisations of the BPS equations summarised in Appendix~\ref{sec:BPSeq}, it is more convenient to work with the gauge-invariant quantities $\omega_F$ and $\tF$, at least when determining solutions which are asymptotically AdS$_3\times S^3$.
We are unable to solve all equations for general mode numbers $(k,m,n)$, however,  we present several explicit examples in the next section and even more in Appendix~\ref{app:Examples}.
We believe that the lack of a general solution is a technical and not a conceptual issue.

\subsection{The zeroth, first and second Layer}

In the supersymmetric ansatz the vector field one-form and its two-form field strength can be decomposed as
\begin{align}
    &A ~ =~ \frac{Z_A}{Z_2}\,(dv + \beta) - \tA\,, && F ~  =~  (dv+ \beta) \wedge \omega_F + \tF\,.
\end{align}
Treating \eqref{eq:PertRkmn} as a perturbation on top of the supertube, so that $Z_2$ is given by \eqref{eq:Supertube}, then we are able to read off%
\footnote{This choice is not unique due to U(1) gauge symmetry, but is convenient to highlight the sources of the fields.}
\begin{subequations}
	\label{eq:PertAnsatz}
	\begin{align}
		Z_A^{(k,m,n)} &\equiv \frac{\sqrt{2}\,\bt\,Q_5}{R_y}\, \frac{\Delta_{k,m,n}}{\Sigma}\,\cos{\hat v_{k,m,n}}\,,\\
		\tA^{(k,m,n)} &\equiv - \bt\, \Delta_{k,m,n}\, \left( \frac{a^2}{r\left(a^2+ r^2\right)}\, dr\, \sin{\hat v_{k,m,n}} - \frac{\sqrt{2}}{R_y}\, \beta_0\,\cos{\hat v_{k,m,n}} \right)\,.
	\end{align}
\end{subequations}
and 
\begin{subequations}
	\label{eq:OmFtF}
	\begin{align}
		\omega_F^{(k,m,n)} &= \frac{\sqrt2\,\bt}{R_y}\,\Delta_{k,m,n}\,\Biggr\lbrace\left[\frac{m\,a^2 + k\,r^2}{r\left(a^2 + r^2\right)}\,dr+\left(\frac{m}{\sin\theta\,\cos\theta}-k\,\cot\theta\right)\,d\theta\right]\cos \hat v_{k,m,n}\nonumber\\*
		&\hspace{4cm}  + \left[(k-m)\,d\phi -m\, d\psi\right]\sin \hat v_{k,m,n}\Biggr\rbrace
		\,,\\
		\tF^{(k,m,n)} &= a^2\,\bt \,\Delta_{k,m,n}\Biggr[ \left(\frac{m}{r\,\sin\theta}-\frac{k\,\cos\theta\,\cot\theta}{r}\right)\,\Omega_1\,\sin \hat v_{k,m,n}\nonumber\\*
		& \hspace{4cm} + \frac{m-k\,\cos^2\theta}{\Sigma}\,\left(\Omega_2+ \Omega_3\right)\,\cos \hat v_{k,m,n} \Biggr]\,,
	\end{align}
\end{subequations}
where we have expanded the two-form in a basis of self-dual two-forms on the base space 
\begin{subequations}
	\label{eq:SDBas}
	\begin{align}
		\Omega_{1} &\equiv \frac{dr\wedge d\theta}{(r^2+a^2)\cos\theta} + \frac{r\sin\theta}{\Sigma} d\phi\wedge d\psi\,,\\
		\Omega_{2} &\equiv  \frac{r}{r^2+a^2} dr\wedge d\psi + \tan\theta\, d\theta\wedge d\phi\,,\\
		\Omega_{3} &\equiv \frac{dr\wedge d\phi}{r} - \cot\theta\, d\theta\wedge d\psi\,,
	\end{align}
\end{subequations}
which manifestly shows that $\tF$ is self-dual on $\IR^4$.
One can show that \eqref{eq:OmFtF}, combined with the supertube backrgound, indeed solve to the BPS equations, in particular \eqref{eq:Lay2}, if one is working to linear order in $\bt$.
At higher order in $\bt$, one needs to take into account that these vector field components act  sources for other fields, which encode the backreaction of the geometry.

Since the differential equations \eqref{eq:Lay2} are linear, any linear combination of solutions  is also a solution.
Hence the most general vector field that solves the BPS equations is given by
\begin{align}
	\omega_F = \sum_{k,m,n}\,\bt_{k,m,n}\, \omega_F^{(k,m,n)} \,,\qquad \tF =  \sum_{k,m,n}\,\bt_{k,m,n}\, \tF^{(k,m,n)}\,.
\end{align}
As already mentioned, these solutions act as sources in the subsequent layers of BPS equations. 
When multiple modes are excited, one obtains source terms associated with the sum and differences of the phases \eqref{eq:vhatPhase}.
In this paper, we limit ourselves to the case where only one mode is excited.
Such solutions, when fully backreacted, are called single-mode superstrata.

From now on, we take the perturbation to be a single-mode solution with mode numbers $(k,m,n)$ --  essentially \eqref{eq:OmFtF}.
The next step is to solve the remaining BPS equations, which gives a solution that is non-linear in $\bt$.
Since the vector field enters only in the second layer BPS equations, we assume that the quantities determined by the zeroth and first layer are unaltered and are those of the round supertube \eqref{eq:Supertube}, in particular
\begin{align}
\dot \beta = \dot h_{mn} = \psi= \dot Z_2 = \Theta^1 = 0.
\end{align}
Exciting only a single mode further simplifies our analysis due a somewhat surprising fact that the quadratic source terms in the third and fourth layers are completely $v$-independent: All terms that ivolve a phase drop out.
Thus there is no need for  ``coiffuring'',  a method in which additional degrees of freedom are excited to remove phase-dependent source terms in the BPS equations which cause singularities in the backreacted metric \cite{Bena:2017xbt}, simply because there are no dangerous terms that need to be cancelled out.

\subsection{Solving the third layer}

The equations in the third layer \eqref{eq:Lay3} determine the quantities $\ttt^2$ and $Z_1$.
We start by noting that $\tF \wedge \tF$ and $\omega_F\wedge \tF$, which act as sources in this layer, are $v$-independent. 
Furthermore, as can be seen in \eqref{eq:Cond2}, $\ttt^2$ and $Z_1$ already contain quadratic contributions from the vector field, which also turn out to be phase-independent.
With that in mind, we assume that $Z_1$ and the fully backreacted $\ttt^2$ are  $v$-independent, which allows us to make an ansatz
\begin{subequations}
	\label{eq:Theta2vindAns}
	\begin{gather}
		\ttt^2 ~=~ d_4\at_2 + \tA \wedge \omega_F + \frac{Z_A}{Z_2}\, \tF  \\
		*_4 d_4 Z_1~=~ d_4 \gamma_1 - \at_2 \wedge d_4\beta- \tF\wedge \tA\,,
	\end{gather}
\end{subequations}
where $\at_2$ and $\gamma_1$ are $v$-independent. 

Since we are working in flat $\IR^4$ with a $v$-independent $\beta$, the BPS equations in this layer simplify to
\begin{subequations}
	\label{eq:ALay1SubLay21vind}
	\begin{gather}
		*_4 \ttt^2   ~=~  \ttt^2 \,,\\
		d_4*_4 d_4 Z_1 + \ttt^2 \wedge d_4 \beta ~=~  \tF\wedge \tF \label{eq:Z1eq}\,,\\
		d_4 \ttt^2 ~=~ - 2\, \omega_F \wedge \tF\,.
	\end{gather}
\end{subequations}
Inserting the ansatz \eqref{eq:Theta2vindAns} solves all equations apart from the self-duality constraint, which reads
\begin{gather}
	\label{eq:a2eq}
	 d_4\at_2 -  *d_4 \at_2 = \frac{\sqrt{2}\,\bt^2}{R_y}\, \Delta_{2k,2m,2n}\left( -\, \frac{m\,a^2}{r^2}\,\oO_2+\frac{(k-m)\,a^2}{a^2+r^2}\oO_3\right) \,.
\end{gather}
The right-hand side is expressed in terms of a basis of anti-self dual two-forms on $\IR^4$,
\begin{subequations}
	\label{eq:asdbasis}
	\begin{align}
		\oO_1 &= \frac{dr\wedge d\theta}{(a^2 + r^2) \cos\theta} - \frac{r \sin\theta}{\Sigma}\, d\phi \wedge d\psi\,,\\
		\oO_2 &= \frac{r}{a^2 + r^2}\, dr\wedge d\psi - \tan\theta d\theta \wedge d\phi\,,\\
		\oO_3 &= \frac{dr \wedge d\phi}{r}+ \cot\theta \, d\theta \wedge d\psi\,,
	\end{align}
\end{subequations}
which satisfy the  following orthogonality conditions
\begin{subequations}
	\begin{align}
		\oO_i \wedge \oO_j &= 0\,, \qquad \text{if} \qquad i \neq j\,,\\
		*^{(0)}\left( \oO_1 \wedge \oO_1\right) &= \frac{*^{(0)}\left( \oO_2 \wedge \oO_2\right)}{\Sigma} = - \frac{2}{(a^2 + r^2) \cos^2 \theta \Sigma^2}\,,\\
		*^{(0)}\left( \oO_3 \wedge \oO_3\right)&= - \frac{2}{r^2 \sin^2 \theta \Sigma}\,,
	\end{align}
\end{subequations}
and are related to \eqref{eq:SDBas} by a change of relative sign between the two terms. 
To solve \eqref{eq:a2eq}, we first note that there is no term proportional to $\oO_1$, so we make an ansatz
\begin{gather}
	\label{eq:a2Ans}
	\at_2 =\frac{\bt^2}{\sqrt2\,R_y}\biggr[ f_1(r,\theta)\left( d\phi + d\psi \right) + f_2 (r,\theta) \left( d\phi - d\psi\right)\biggr]\,.
\end{gather}
Inserting this into \eqref{eq:a2eq} yields two first-order differential equations for $f_1(r,\theta)$ and $f_2(r,\theta)$.
\begin{subequations}
	\label{eq:oocomp}
	\begin{gather}
		\frac{a^2 + r^2}{2\,r^2}\, \left(\pd_r f_1 - \pd_r f_2\right) - \frac{\cot\theta}{2}\left(\pd_\theta f_1 + \pd_\theta f_2\right) = -  \Delta_{2k,2m,2n}\, \frac{m\,a^2}{r^2}\,,\\
		\frac{r}{2}\, \left(\pd_r f_1 + \pd_r f_2\right) + \frac{\tan\theta}{2}\left(\pd_\theta f_1 - \pd_\theta f_2\right) =  \Delta_{2k,2m,2n}\, \frac{(k-m)\,a^2}{a^2+r^2}\,,
	\end{gather}
\end{subequations}
which can be used to express the derivatives of $f_1$ only in terms of $f_2$
\begin{subequations}
	\label{eq:pdf1}
	\begin{align}
		\pd_r f_1 &= \frac{2\,r\,\tan\theta}{r^2 + (a^2 + r^2)\tan^2\theta}\,\pd_\theta f_2 + \left(1- \frac{2\,r^2}{r^2 + (a^2 + r^2)\tan^2\theta}\right)\,\pd_rf_2 \nonumber\\
		& \quad + \frac{\Delta_{2k,2m,2n}}{r^2 + a^2 \sin^2\theta}\left(\frac{2a^2\, (k-m) \,r \cos^2 \theta}{a^2 + r^2}- \frac{2 a^2 \,m \,\sin^2\theta}{r}\right)\,,
		\\
		\pd_\theta f_1 &= \left(1- \frac{2\,r^2}{r^2 + (a^2 + r^2)\tan^2\theta}\right)\,\pd_\theta f_2 - \frac{2\,r\,(a^2 + r^2)\,\tan\theta}{r^2 + (a^2 + r^2)\tan^2\theta}\,\pd_r f_2
		\nonumber\\
		& \quad + \frac{\Delta_{2k,2m,2n}}{r^2 + a^2 \sin^2\theta}\,2 a^2\,k\, \sin\theta \, \cos\theta\,.
	\end{align}
\end{subequations}
Next, one either differentiates these equations with respect to $r$ and $\theta$ or acts with an exterior derivative on \eqref{eq:a2eq} to obtain four new equations. 
Three of those are used to express all second derivatives of $f_1$ in terms of $f_2$.
What remains is a Laplace equation on the base space determining $f_2$
\begin{align}
	\label{eq:Lap1}
	\cLh f_2 &= -\frac{2\,a^2}{\Sigma(a^2 + r^2)\, \cos^2\theta}\,\Big((k-m)^2\, \Delta_{2k,2m+2,2n} + m^2\, \Delta_{2k,2m,2n-2}\nonumber\\* &\quad  -k\,(m+n)\, \Delta_{2k+2, 2m+2, 2n-2}\Big)\,,
\end{align}
where $\cLh$ is the scalar Laplace operator on the 4-dimensional base space
\begin{align}
\label{eq:LapDef}
	\cLh f(r,\theta) \equiv - *_4 d_4 *_4 d_4\,f =\frac{1}{r\, \Sigma}\,\pd_r \left(r\left(a^2 + r^2\right)\pd_r f\right)+ \frac{1}{\Sigma\, \cos\theta\, \sin\theta}\,\pd_\theta\left(\cos\theta\, \sin\theta\pd_\theta f\right)\,.
\end{align}
We now use the fact that the differential equation of the form
\begin{align}\label{solvableeq}
	\widehat{\cL}F_{2k,2m,2n}={\Delta_{2k,2m,2n}\over (r^2+a^2)\cos^2\theta\,\, \Sigma} 
\end{align}
is solved by the function \cite{Bena:2015bea,Bena:2017xbt}
\begin{align} \label{Ffun}
	F_{2k,2m,2n}&=-\!\sum^{j_1+j_2+j_3\le k+n-1}_{j_1,j_2,j_3=0}\!\!{j_1+j_2+j_3 \choose j_1,j_2,j_3}\frac{{k+n-j_1-j_2-j_3-1 \choose k-m-j_1,m-j_2-1,n-j_3}^2}{{k+n-1 \choose k-m,m-1,n}^2}\notag\\
	&\qquad\qquad\qquad\qquad\qquad\qquad
	\times
	\frac{\Delta_{2(k-j_1-j_2-1),2(m-j_2-1),2(n-j_3)}}{4(k+n)^2(r^2+a^2)}\,,
\end{align} 
with 
\begin{equation} 
	{j_1+j_2+j_3 \choose j_1,j_2,j_3}\equiv \frac{(j_1+j_2+j_3)!}{j_1!\, j_2!\, j_3!}\,.
\end{equation} 
Since the differential equation \eqref{eq:Lap1} is already in the required form, we find that the solution is given by
\begin{align}
	\label{eq:f2sol}
	f_2^{(k,m,n)}(r,\theta) &= -2 \,a^2\Big[(k-m)^2\, F_{2k,2m+2,2n} + m^2\, F_{2k,2m,2n-2}   - k\,(m+n)\, F_{2k+2, 2m+2, 2n-2} \Big]\,.
\end{align}
Next one obtains $f_1$  by integrating the differential equations \eqref{eq:pdf1}.  While we are not able to find a closed-form expression for this function, it is easily determined  case-by-case.
With $f_1$ and $f_2$ known, one can then calculate $\at_2$ and $\ttt^2$ for all mode numbers, 
and check that they indeed solve all third layer BPS equations \eqref{eq:ALay1SubLay21vind}. 

Interestingly, all BPS equations can be solved without explicitly calculating $Z_1$.%
\footnote{In the third layer, $Z_1$ only appears through $d_4*_4d_4Z_1$,  so that one only needs to know $\at_2$. In the fourth layer, $Z_1$ always multiplies a vanishing quantity, so it never directly appears in the equations.}
However, its precise form is crucial for the properties of the six-dimensional  metric. 
To determine $Z_1$, we rearrange \eqref{eq:Z1eq} to
\begin{align}
\label{eq:Z1Lap}
    \cLh Z_1 =  *_4\left(\ttt^2 \wedge d_4 \beta - \tF \wedge\tF\right)\,.
\end{align}
It is important to highlight that the right-hand side contains two terms which contribute with opposite sign. 
The term involving $\ttt^2$ is also present in the superstrata built with tensor multiplets and typically gives a positive contribution.
The second term comes solely from the vector-field excitation and contributes with opposite sign, in essence decreasing the charge of $Z_1$. 
To be precise, we can find
\begin{subequations}
    \begin{align}
     *_4\ttt^2 \wedge d_4 \beta &=  \frac{4\,a^2\, \bt^2}{\Sigma^3\,(r^2+ a^2
\sin^2\theta)}\left(\,a^2 \,\sin\theta\,\cos\theta \,\pd_{\theta} f_2 -  r(a^2 + r^2) \pd_r f_2 \right)\nonumber\\
&\quad +\frac{4 \,a^4\,\bt^2\,k}{\Sigma^3\,(r^2+ a^2
\sin^2\theta)}\,\Delta_{2k,2m,2n}\label{eq:Z1term1}\,,\\
    *_4 \tF\wedge \tF &= \frac{2\,a^2\,\bt^2}{\Sigma^2}\, \frac{\Delta_{2k+2, 2m, 2n-2}}{(a^2 + r^2)\, \cos^2\theta\,\sin^4\theta}\,\left(k \, \cos^2\theta -m\right)^2\,. 
\end{align}
\end{subequations}
Notice that the expressions involve divergences at the location of the supertube, where $\Sigma\rightarrow 0$, however, this merely signals that $Z_1$ goes as $\Sigma^{-1}$.
This can be also seen by the fact that the differential operator acting on $f_2$ in \eqref{eq:Z1term1} is given by the difference of two Laplace operators
\begin{align}
	\label{eq:DiffOper}
    	\cL_{\Sigma}f(r,\theta) \equiv \frac{4}{\Sigma^3}\left(a^2 \,\sin\theta\,\cos\theta \pd_{\theta}  -  r(a^2 + r^2) \pd_r \right)f(r,\theta) = \cLh\left(\frac{f(r,\theta)}{\Sigma}\right) -\frac{1}{\Sigma}\, \cLh f(r,\theta) \,.
\end{align}
Despite the fact that $f_2$ is known explicitly, we are not able to find a closed form expression for \eqref{eq:Z1term1}. 
Nonetheless, the right-hand side of \eqref{eq:Z1Lap} can be computed on a case-by-case basis for arbitrary mode numbers and in all cases we analysed we were able to solve the differential equation.
We present some of the explicit solutions in Section~\ref{sec:ExpExam}.

\subsection{Solving the fourth Layer}

In principle, the BPS equations determining $\cft$ and $\omega$ are coupled and need to be solved simultaneously. 
However, given the solution of the previous layers, only a single term on the right-hand side of \eqref{eq:Lay4Eq1} is non-vanishing
\begin{align}
	Z_2 \, \omega_F^2 = \frac{2 \,\bt^2\, Q_5}{R_y^2}\, \frac{1}{{\Sigma\,(a^2 + r^2)\, \cos^2\theta}}\left[(k-m)^2\,\Delta_{2k,2m+2,2n} + m^2\, \Delta_{2k,2m,2n-2}\right]\,.
\end{align}
As it is $v$-independent, we again make the assumption that $\omega$ and $\cft$ are $v$-independent. 
Therefore the equation \eqref{eq:Lay4Eq1} reduces to
\begin{align}
    \cLh\,\cft =\frac{4 \,\bt^2\, Q_5}{R_y^2}\, \frac{1}{{\Sigma\,(a^2 + r^2)\, \cos^2\theta}}\left[(k-m)^2\,\Delta_{2k,2m+2,2n} + m^2\, \Delta_{2k,2m,2n-2}\right]\,,
\end{align}
which is solved by
\begin{align}
	\label{eq:cFsol}
	\cft^{(k,m,n)} =  \frac{4\, \bt^2\, Q_5}{R_y^2}\left[(k-m)^2\,F_{2k,2m+2,2n} + m^2\, F_{2k,2m,2n-2}\right]\,.
\end{align}
Interestingly, this results is, up to constant factors, exactly  the first two terms appearing in the expression for $f_2$ \eqref{eq:f2sol}.
We find it convenient  extract the prefactors and define
\begin{align}
\label{eq:fDef}
    \cft \equiv  \frac{4\, \bt^2\, Q_5}{R_y^2}\,f(r,\theta)\,.
\end{align}

The last unsolved equation determines $\omega$
\begin{align}
    	\label{eq:Omegaeqkmn}
	d_4 \omega + *_4 d_4\omega &= Z_2 \, \ttt^2 - \cft \, d_4\beta = \frac{\sqrt{2}\,\bt^2\,Q_5}{R_y}\left(g_{2}\,\Omega_2 + g_3\,\Omega_3 \right)\,,
\end{align}
where we expanded in the self-dual basis \eqref{eq:SDBas} and
\begin{subequations}
    \begin{align}
        g_2 &\equiv \frac{a^2\,k\,\cos^2\theta\,\Delta_{2k,2m,2n}}{\Sigma\left(r^2 + a^2\,\sin^2\theta\right)} + \frac{(a^2 + r^2) \,\sin\theta\,\cos\theta}{\Sigma\left(r^2 + a^2\,\sin^2\theta\right)}\,\pd_\theta f_2(r,\theta)- \frac{(a^2 + r^2) \,r\,\cos^2\theta}{\Sigma\left(r^2 + a^2\,\sin^2\theta\right)}\,\pd_r f_2(r,\theta)\nonumber\\*
        & \quad - \frac{4\,a^2\,(a^2 + r^2) \,\cos^2\theta}{\Sigma^2}\,f(r,\theta) \,,\\
        g_3 &\equiv -\frac{a^2\,k\,\sin^2\theta\,\Delta_{2k,2m,2n}}{\Sigma\left(r^2 + a^2\,\sin^2\theta\right)} + \frac{r^2\,\sin\theta\,\cos\theta}{\Sigma\left(r^2 + a^2\,\sin^2\theta\right)}\,\pd_\theta f_2(r,\theta)+ \frac{(a^2 + r^2) \,r\,\sin^2\theta}{\Sigma\left(r^2 + a^2\,\sin^2\theta\right)}\,\pd_r f_2(r,\theta)\nonumber\\*
        & \quad + \frac{4\,a^2\,r^2 \,\sin^2\theta}{\Sigma^2}\,f(r,\theta) \,.
    \end{align}
\end{subequations}
We make the following ansatz for the $\bt^2$ contribution to $\omega$ 
\begin{align}
\label{eq:OmegaBAns}
	\omega_{\bt} = \frac{\sqrt2\,Q_5\,\bt^2}{R_y} \Big[\mu(r,\theta)\, \left( d\phi + d\psi \right) + \nu (r,\theta) \left( d\phi - d\psi\right)\Big]\,,
\end{align}
which inserted into \eqref{eq:Omegaeqkmn} yields two independent equations that can be used to derive
\begin{subequations}
\label{eq:nuEq}
	\begin{align}
		\pd_r\nu &= \frac{2\,a^2\,k\,r\,\sin^2\theta\,\cos^2\theta }{\Sigma\,\left(r^2 + a^2\,\sin^2\theta\right)^2}\Delta_{2k,2m,2n} -\frac{2\,r\,\sin\theta\,\cos\theta}{r^2 + a^2\,\sin^2\theta}\,\pd_\theta\mu- \frac{\left(a^2 + r^2 - r^2\,\cot^2\theta\right)\,\sin^2\theta}{r^2 + a^2\,\sin^2\theta}\,\pd_r\mu\nonumber\\* 
		& \quad - \frac{2\,r^2\,(a^2 + r^2)\, \sin^2\theta\,\cos^2\theta }{\Sigma\,\left(r^2 + a^2\,\sin^2\theta\right)^2}\,\pd_r  
		f_2
		+ \frac{r\,\left(a^2 - (a^2 + 2r^2)\cos2\theta\right)\,\sin\theta\,\cos\theta }{2\,\Sigma\,\left(r^2 + a^2\,\sin^2\theta\right)^2}\,\pd_\theta 	f_2\nonumber\\* 
		& \quad
		- \frac{4\,a^2\,r\,(a^2 + 2r^2)\,\sin^2\theta\,\cos^2\theta }{\Sigma^2\,\left(r^2 + a^2\,\sin^2\theta\right)}\,f\,,
		\\
		\pd_\theta \nu &=  \frac{a^2\,k\,\left(a^2 - (a^2 + 2r^2)\cos2\theta\right)\,\sin^2\theta\,\cos^2\theta }{\Sigma\,\left(r^2 + a^2\,\sin^2\theta\right)^2}\Delta_{2k,2m,2n} 
		+ \frac{r^2 \,\cos2\theta- a^2\,\sin^2\theta}{r^2 + a^2\,\sin^2}\,\pd_\theta\mu  \nonumber\\
		& \quad 
		+ \frac{2\,r \,(a^2 + r^2) \,\sin\theta\,\cos\theta}{r^2 + a^2\,\sin^2}\,\pd_r\mu 
		- \frac{2\,r^2\,(a^2 + r^2)\, \sin^2\theta\,\cos^2\theta }{\Sigma\,\left(r^2 + a^2\,\sin^2\theta\right)^2}\,\pd_\theta 
		f_2\nonumber\\* 
		& \quad
		- \frac{r\,\left(a^2 - (a^2 + 2r^2)\cos2\theta\right)\,\sin\theta\,\cos\theta }{2\,\Sigma\,\left(r^2 + a^2\,\sin^2\theta\right)^2}\,\pd_r 	f_2
		+ \frac{a^2\,r^2\,(a^2 + 2r^2)\,\sin4\theta}{\Sigma^2\,\left(r^2 + a^2\,\sin^2\theta\right)}\,f
		\,.
	\end{align}
\end{subequations}
One can then differentiate these expressions once more and obtain equations for the second derivatives of $\nu$. 
After some algebra, one is left with a Laplace equation for $\mu$ 
\begin{align}
    \label{eq:muEq}
 &\cLh\left(\mu + \frac{r^2 + a^2 \,\sin^2\theta}{2\,\Sigma}\,f\right) = - \frac{a^2}{4 \,\left(r^2 + a^2 \,\sin^2\theta\right)}\,\cL_{\Sigma}f_2\nonumber\\
 &+ \frac{r^2+ a^2 \sin^2\theta}{2\,\Sigma^2\,(a^2 + r^2)\, \cos^2\theta}\left[(k-m)^2\,\Delta_{2k,2m+2,2n} + m^2\, \Delta_{2k,2m,2n-2}\right] \nonumber\\*
 & -\frac{1}{\Sigma\,\left(r^2+ a^2 \sin^2\theta\right)\,\cos^2\theta}\,\left[(k-m)^2\,\Delta_{2k,2m+2,2n+2} + m^2\, \Delta_{2k,2m,2n-2}\right]\nonumber\\
 & + \frac{\Delta_{2k,2m,2n}}{\Sigma\,\left(r^2+ a^2 \sin^2\theta\right)}\left[\frac{a^4(k-m)m}{r^2\,(a^2 + r^2)} - \frac{a^4\,k}{\Sigma^2} - \frac{a^2\,(k-2m)(a^2\,m+k\,r^2)\,(a^2 + 2r^2)}{r^2\,(a^2 + r^2)\,\Sigma}\right]\,.
\end{align}
Again, we are not able to find a closed-form solution for all mode-numbers.
However, we are able to solve this equation on a case-by-case basis for any mode number.
Once $\mu$ is found,  one can determine $\nu$ by integrating \eqref{eq:nuEq}.

\section{Explicit examples}
\label{sec:ExpExam}

In this section we present the full solutions of superstrata with mode numbers $(1,0,n)$ and $(1,1,n)$ and analyse their properties. 
Several other solutions, including the full expressions for all the field content of  $(1,0,1) $ and $(1,1,1)$ superstrata, are  presented in Appendix~\ref{app:Examples}.
By examining geometries with generic mode numbers, we determine the general expressions for momentum $Q_P$ and angular momenta $J$ and $\Jb$ and find complete agreement with the values predicted on the CFT side.
We also comment on the extension of these solutions to linear dilaton and  asymptotically flat regions.

For convenience we recall that we made the following ansatze
\begin{gather}
\at_2 =\frac{\bt^2}{\sqrt2\,R_y}\biggr[ f_1(r,\theta)\left( d\phi + d\psi \right) + f_2 (r,\theta) \left( d\phi - d\psi\right)\biggr]\,,\\
    \cft \equiv  \frac{4\, \bt^2\, Q_5}{R_y^2}\,f(r,\theta)\,,\qquad \omega_{\bt} = \frac{\sqrt2\,Q_5\,\bt^2}{R_y} \Big[\mu(r,\theta)\, \left( d\phi + d\psi \right) + \nu (r,\theta) \left( d\phi - d\psi\right)\Big]\,,
\end{gather}
which we use to present the full solutions.

\subsection{$(k,m,n) = (1,0,n)$}

We first analyse the solution with mode numbers $(k,m,n)=(1,0,n)$.
The various quantities parametrising the solutions are given by
\begin{subequations}
	\label{eq:10nFamily}
	\begin{align}
         Z_1^{(1,0,n)} &= \frac{Q_1}{\Sigma}- \frac{\bt^2 }{2\,n\,(n+1)^2\Sigma}\left[1- \frac{r^2+ (2n+1)a^2 \, \cos^2\theta}{n \, a^2}\left(1- \Delta_{0,0,2n}\right)+ n\, \Delta_{2,2,2n}\right]\,,\\
		  f^{(1,0,n)} &= -\frac{1}{4\,(n+1)^2\, a^2}\,\left(1 - \Delta_{0,0,2n+2}\right) \,,\\
		f^{(1,0,n)}_1&= - \frac{1}{2\,n\,(n+1)^2}\,\left[\frac{r^2 + a^2\, \sin^2\theta}{a^2}\left(1-\Delta_{0,0,2n}\right)-n\,\Delta_{0,0,2n+2}-n\,(n+2)\,\Delta_{2,0,2n}\right]\,,\\
		f_2^{(1,0,n)} &= \frac{1}{2(n+1)^2}\,\left(1 - \Delta_{0,0,2n+2}\right)- \frac{1}{2\,n\,(n+1)^2}\Biggr[-n^2\,\Delta_{2,0,2n}-2\,n\,\sin^2\theta\, \Delta_{0,0,2n}\nonumber\\& \quad + \frac{a^2+ 2r^2}{a^2}\,\sin^2\theta\left(1-\Delta_{0,0,2n}\right)   - \frac{r^2}{a^2}\left(1-\Delta_{0,0,2n}\right)\Biggr]\,\\
		\mu^{(1,0,n)} &= - \frac{r^2 + a^2\, \sin^2\theta}{\Sigma}\, f^{(1,0,n)} - \frac{1}{4\,n\,(n+1)\,\Sigma}\frac{r^2}{a^2}\left(1-\Delta_{0,0,2n}\right)\,\\
		\nu^{(1,0,n)} &= -\frac{1}{4\,(n+1)^2\,\Sigma}\left[1+\frac{r^2\cos(2\theta)}{n\,a^2}\,\left(1-\Delta_{0,0,2n}\right)- \cos^2\theta\left(1+ \Delta_{0,0,2n+2}\right)\right]\,,\\
		\gamma_1^{(1,0,n)} &=\frac{\cos^2\theta}{\Sigma}\left[-  Q_{1}\,(r^2 + a^2) + \frac{\bt^2}{2\,n\,(n+1)}\,\Delta_{2,0,2n}\left((n+1)\,a^2 + r^2\right)\right]d\phi \wedge d\psi \,. 
	\end{align}
\end{subequations}
It is important to note that $Z_1$ and $\mu$ one can have additional homogeneous terms, which we have fixed so that the metric is smooth at the origin and the asymptotic radii of AdS$_3$ and $S^3$ are $R_{\rm AdS}^2 = \sqrt{Q_1\,Q_5}$.

\begin{figure}[t]
	\includegraphics[width=\textwidth]{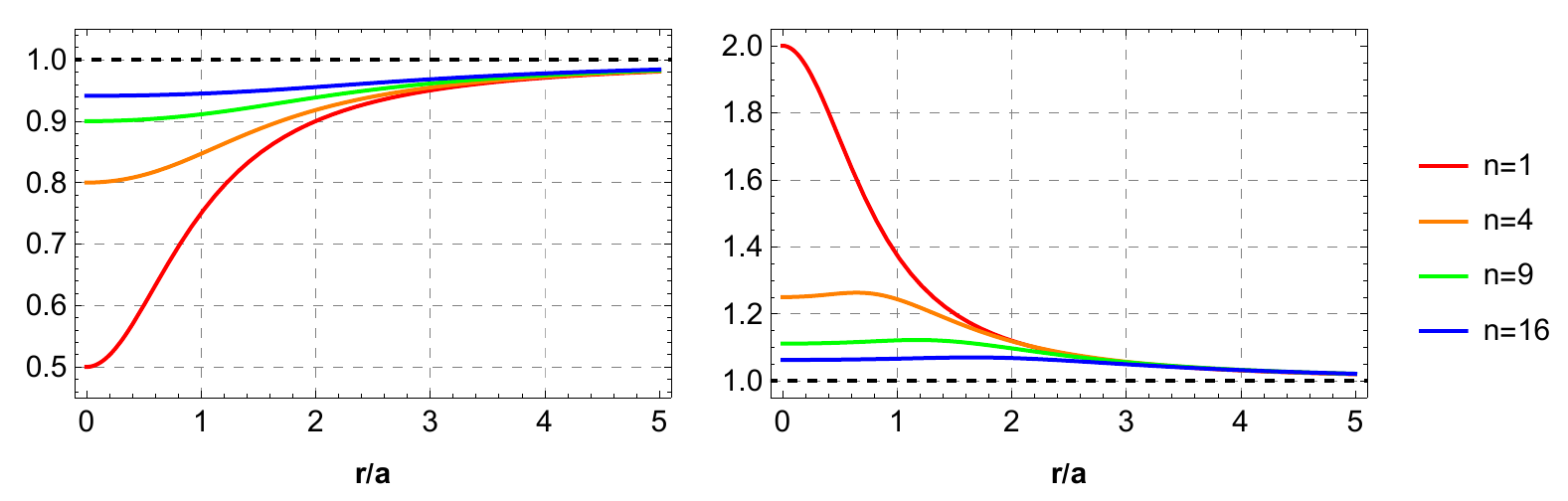}
	\label{fig:Z1Plot10n}
 \vspace{-0.5cm}
	\caption{The ratio between  $Z_1^{(1,0,n)}$ and $Z_1^{(1,0,0)} = \tfrac{Q_1}{\Sigma}$ as a function of the radial coordinate.
	We have taken $Q_1 = Q_5 = 10^{10}$, $R_y = 10^7$ and $a=1$.
	On the left  $\theta = \tfrac{\pi}2$, while on the right $\theta = 0$.
    From the former we can  can extract the behaviour near the brane sources at $r \to 0$. As the radial distance increases, we find a monotonically increasing function for all solutions, indicating that there is additional $Q_1$ charge dissolved in the geometry, which adds to the charge  as one approaches the asymptotic region.}
\end{figure}

The behaviour of the  $Z_1$ as a function of $r$ is given in Figure~\ref{fig:Z1Plot10n}, where we plot it for two values of $\theta$. 
From the divergence at $\Sigma \to 0$ we know that this harmonic function is sourced by charges are localised on the supertube.
This behaviour can be isolated by introducing 
\begin{align}
    r = a\,\lambda\, \cos\chi\,, \qquad \theta = \frac{\pi}{2}- \lambda\,\sin\chi\,,\label{eq:STLim}
\end{align}
followed by taking $\lambda \to 0$. 
We find that 
\begin{align}
\label{eq:Z110nLocal}
    Z_1^{(1,0,n)} \xrightarrow[\lambda \to 0]{} \frac{1}{a^2\,\lambda^2}\,\left(Q_1-\frac{\bt^2}{2\,n\,(n+1)^2}\right)\,.
\end{align}
On the other hand, near the boundary of AdS  we find
\begin{align}
	\label{eq:AsymptoticCharge}
    Z_1 \xrightarrow[r\to \infty]{}\frac{Q_1}{r^2}\,.
\end{align}
We see that in some way there is more $Q_1$ charge dissolved in the geometry  which gets added up to the total charge (see Figure~\ref{fig:Z1Plot10n}).%
\footnote{Note that the regularity condition \eqref{eq:Regularitybt}, which ensures that the solution is smooth, prevents the parenthesis in \eqref{eq:Z110nLocal} to ever become negative. This can be also seen on the left in Figure~\ref{fig:Z1Plot10n}.}

\begin{figure}[t]
	\includegraphics[width=\textwidth]{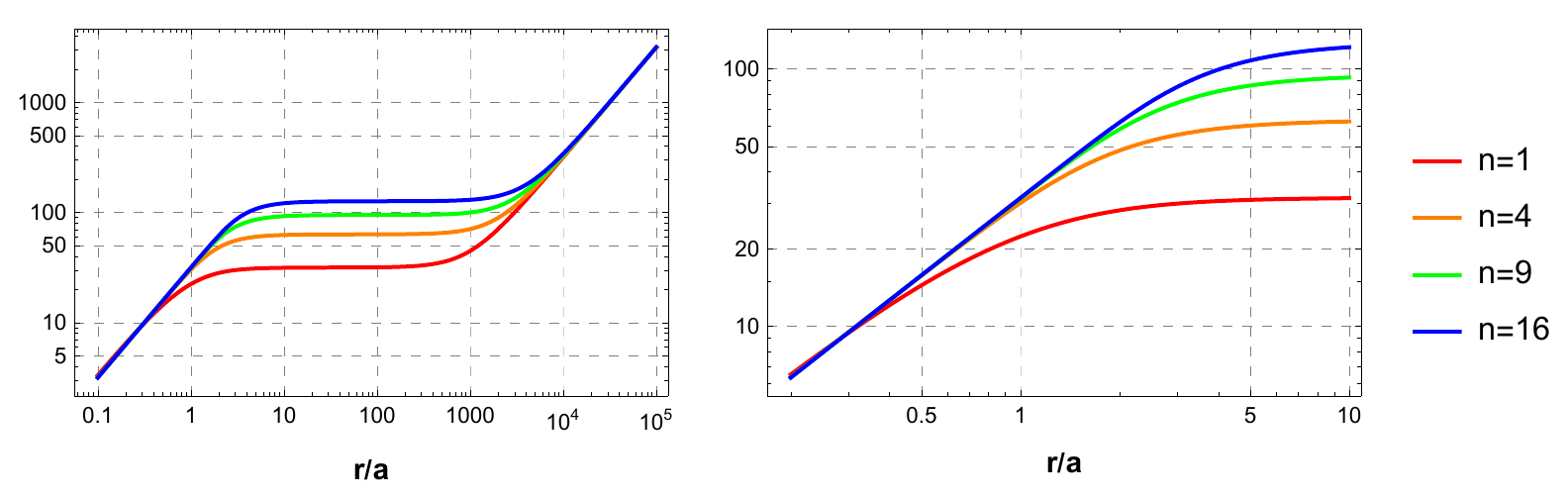}
	\label{fig:10n}
  \vspace{-0.5cm}
	\caption{The log-log plot of $\sqrt{f_y}$ as a function of the radial coordinate $r$ for several members of the $(k,m,n) = (1,0,n)$ family.
	We have taken $\theta = \tfrac{\pi}4$ and $Q_1 = Q_5 = 10^{10}$, $R_y = 10^7$ and $a=1$.
	Linear growth characterises AdS$_3$ regions while the constant parts indicate the stabilisation of the size of the $S^1_y$-circle and the onset of an AdS$_2$ throat. 
	On the right plot we show the details of the near-source region.		
}
\end{figure}

Next we can rewrite the six-dimensional metric in the following fibered form
\begin{align}
    ds^2_6 =  f_{r}\, dr^2 + f_{t}\, dt^2 + f_{y}\,\left(dy+ A^{(y)}\right)^2+ f_{\theta}\,\,d\theta^2 +
    f_{\phi}\,\left( d\phi + A^{(\phi)}\right)^2 + f_{\psi}\,\left( d\psi + A^{(\psi)}\right)^2\,,
    	\label{eq:FiberedMetric}
\end{align}
where the $A^{(M)}$ are one-forms on  AdS$_3$. 
The expressions for these one-forms and functions $f_{M}$ are tedious for generic $n$, so we do not give them explicitly here. 
By analysing the coefficients $f_M$, one can check that, subject to the regularity condition \eqref{eq:Regularity}, no compact direction diverges at either the origin of the base space ($r=0$ and $\theta = 0$) or at the location of the supertube ($\Sigma = 0$, or $r=0$, $\theta = \pi/2$).
In particular, in Figure~\ref{fig:10n}  we plot the behaviour of  $f_y$, which contains the information about the size of the $S^1_y$-circle.
We see that the geometries have three different regions: They are asymptotically AdS$_3$, as characterised by the linear growth at large values of $r$.
They then develop an AdS$_2\times S_y^1$ throat, in which the radius of the circle stabilises. 
This throat is then smoothly capped off at $r \sim a$ with another AdS$_3$ region. 
Vector superstrata thus behave exactly the same as the previously constructed solutions \cite{Bena:2015bea, Bena:2016ypk, Bena:2017xbt, Ceplak:2018pws}.

\subsection{$(k,m,n) = (1,1,n)$}

The explicit solution for this family is given by the following functions
\begin{subequations}
	\label{eq:11nFamily}
	\begin{align}
		Z_1^{(1,1,n)} &= \frac{Q_1}{\Sigma}+ \frac{\bt^2}{2n^2\,(n+1)\,\Sigma}\Biggr[1+\frac{1}{n+1}\biggr(\frac{r^2}{a^2}\left(1-\Delta_{0,0,2n-2}\right) -2n \Delta_{0,0,2n} - n^2\, \Delta_{2,0,2n}\nonumber\\*
		&\quad \hspace{10ex}-(2n+1)\cos^2\theta\left(1-\Delta_{0,0,2n}\right)\biggr)\Biggr]\,,\\
         f^{(1,1,n)} &= -\frac{1}{4\,n^2\, a^2}\,\left(1 - \Delta_{0,0,2n}\right) \,,\\
		f^{(1,1,n)}_1&=  \frac{1}{2\,n^2\,(n+1)}\,\left[\frac{r^2 + a^2\,\sin^2\theta}{a^2}\left(1-\Delta_{0,0,2n}\right)-n\,\Delta_{0,0,2n}-n^2\,\Delta_{2,2,2n}\right]\,,\\
		f_2^{(1,1,n)} &= -\frac{1}{2(n+1)\,n^2}\,\Biggr[-n^2 \Delta_{2,2,2n} - n \, \cos(2\theta)\Delta_{0,0,2n}+ \frac{\left(a^2 + 2r^2\right)\cos^2\theta-a^2-r^2}{a^2}\nonumber\\*&\quad \hspace{10ex}\times \left(1-\Delta_{0,0,2n}\right)\Biggr]\,\\
		\mu^{(1,1,n)} &= \frac{r^2+ (n+1)\,a^2\,\sin^2\theta}{4\,n^2(n+1)\,a^2\, \Sigma} \left(1- \Delta_{0,0,2n} \right)\,\\
		\nu^{(1,1,n)} &= \frac{1}{4\,n^2\,\Sigma}\Biggr[-1+\frac{1}{n+1}\biggr(\frac{r^2\cos(2\theta)}{a^2}\,\left(1-\Delta_{0,0,2n-2}\right)+(n+2\cos^2\theta)\Delta_{0,0,2n}\nonumber\\*&\quad \hspace{10ex}+(n+1)\cos^2\theta\left(1-\Delta_{0,0,2n}\right)\biggr)\Biggr]\,,\\
		\gamma_1^{(1,1,n)} &=\frac{\cos^2\theta}{\Sigma}\left[-  Q_{1}\,(r^2 + a^2) - \frac{\bt^2}{2\,n\,(n+1)}\,\Delta_{2,0,2n}\left((n+1)\,a^2 + r^2\right)\right]d\phi \wedge d\psi \,. 
	\end{align}
\end{subequations}

\begin{figure}[t]
	\includegraphics[width=\textwidth]{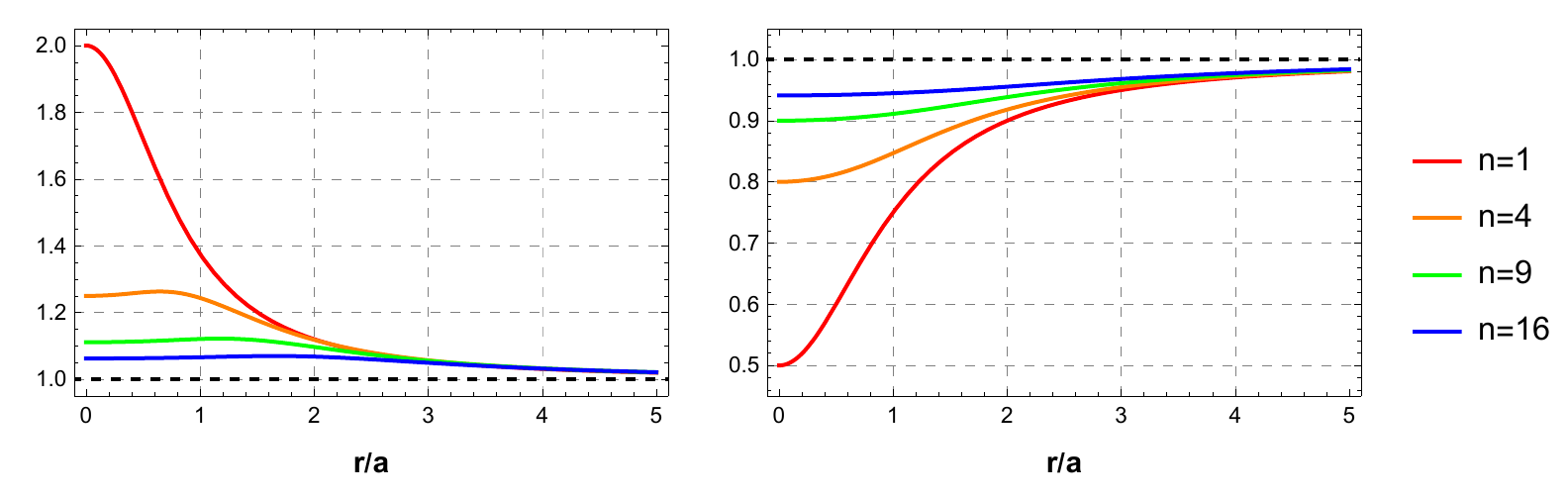}
	\label{fig:Z1Plot11n}
  \vspace{-0.5cm}
	\caption{The plot of the ratio between $Z_1^{(1,1,n)}$ and $Z_1^{(1,0,0)} = \tfrac{Q_1}{\Sigma}$ as a function of the radial coordinate.
	In the plots, we have taken $Q_1 = Q_5 = 10^{10}$, $R_y = 10^7$ and $a=1$. On the left $\theta = \tfrac{\pi}{2}$ and on the right $\theta = 0$.
	The left plot indicates that the charge near the branes is larger than at infinity.}
\end{figure}

Most properties of these superstrata are similar to those of the $(1,0,n)$ family, in particular, the size of the $y$-circle exhibits the same AdS$_3$-AdS$_2$-AdS$_3$ transitions as depicted in Figure~\ref{fig:10n}.
There is a difference in the behaviour of $Z_1$ functions. 
While their values match asymptotically, \eqref{eq:AsymptoticCharge}, one finds that near the supertube location
\begin{align}
\label{eq:Z11nLocal}
    Z_1^{(1,1,n)} \xrightarrow[\lambda \to 0]{} \frac{1}{a^2\,\lambda^2}\,\left(Q_1+\frac{\bt^2}{2\,n^2\,(n+1)}\right)\,.
\end{align}
This indicates that there is more  charge near the brane sources than asymptotically: There seems to be negative charge distributed in the bulk of the geometry (see Figure~\ref{fig:Z1Plot11n}).

\subsection{Properties of solutions with arbitrary mode numbers}

While at this point we are unable to provide explicit expressions for a single-mode solutions with arbitrary mode numbers $(k,m,n)$, we can deduce some general their general properties.

\subsubsection{Regularity}

Generically, one can add undetermined homogeneous terms to $\mu$ and $Z_1$m in particular, one free to add terms that scale as $\Sigma^{-1}$.
We choose to fix such terms in $Z_1$  by demanding that at large distance this harmonic function behaves as
\begin{align}
	Z_1 \xrightarrow{r\to \infty} \frac{Q_1}{r^2}+ \cO\left(r^{-3}\right)\,,
\end{align}
as ensures that the asymptotic radius of AdS$_3$ is given by $R_{\rm AdS}^2= \sqrt{Q_1\, Q_5}$.
As a consequence, the $Z_1$ charge near the supertube locus is different than the charge measured at infinity. 
Whether it increases or decreases with radial distance depends on the value of $m$.
For $m=0$ the charge increases, suggesting that there is more positive charge dissipated in the geometry. 
On the other hand, when $m >0$,  the charge decreases with the distance from the supertube, exactly in the same manner as we saw for the $(1,1,n)$ superstrata.
This suggests that there is charge of opposite sign, compared to the supertube, which is distributed in the geometry. 
One might expect  such solutions to be incompatible with supersymmetry, however, we find no violation of the BPS equations nor any appearance of closed timelike curves.   
Since $m \propto j - \jb$, one may wonder whether non-equal rotation in the $S^3$ induces a kind of screening effect and which effectively decreases measured $Z_1$ charge away from the supertube location.

The homogeneous term in $\mu$ is determined by looking at the behaviour of the $(d\phi+ d\psi)^2$ and $(d\phi- d\psi)^2$ components of the metric near the origin of $\IR^4$, where $r \to 0$ and $\theta \to 0$.
The metric is smooth provided
\begin{align}
\label{eq:Regularitybt}
	Q_1\, Q_5 = a^2\, R_y^2 + x_{k,m,n}\, \frac{\bt^2\,Q_5}{2}\,,
\end{align} 
where we introduced a numerical factor 
\begin{align}
	x_{k,m,n} \equiv k\,\frac{m!\,(n-1)!\,(k-m)!}{(k+n)!}\,.
\end{align}
Recall that $\bt$ is a dimensionful parameter that regulates the magnitude of the vector perturbation.
However, introducing a rescaled parameter
\begin{align}
\label{eq:btobt}
	b^2 \equiv \bt^2\, \frac{Q_5}{R_y^2}\,,
\end{align}
 seems to be more appropriate as in this case the regularity condition takes on a more familiar form 
 \begin{align}
	\label{eq:Regularity}
	Q_1\, Q_5 = a^2\, R_y^2 + \frac{x_{k,m,n}}{2}\,b^2\,R_y^2\,.
\end{align} 
The rescaling \eqref{eq:btobt} shows that the strength of the backreaction is not only influenced by the magnitude of the perturbation $\bt$, but also by the ratio between $Q_5$ and $R_y$, which in principle this allows for regulating non-linear perturbations by modifying the moduli of the theory.

\subsubsection{Coupling to flat space}

The geometries constructed via the method described in the previous sections are asymptotically AdS$_3\times S^3$. 
To couple them to flat space, one has to reinstate the factors of unity in the harmonic functions
\begin{align}
\label{eq:AsyFlat}
    Z_{1} \xrightarrow[]{} 1+ Z_{1}\,, \qquad  Z_{2} \xrightarrow[]{} 1+ Z_{2}\,.
\end{align}
In the original superstrata \cite{Bena:2015bea, Bena:2016ypk, Bena:2017xbt}, this introduces $v$-dependent source terms in the BPS equations.%
\footnote{Interestingly, for supercharged superstrata the coupling to flat space through \eqref{eq:AsyFlat} is trivial \cite{Ceplak:2018pws}.}
In the case of vector fields, extending the geometries to flat space is slightly more complicated.
This is due to  $Z_2$ appearing in the denominator of the U(1) gauge-field decomposition \eqref{eq:Potentials}, in which case adding a constant factor adds non-trivial source terms in the fourth layer. 

In fact, there already exist some examples of non-trivial geometries in asymptotically flat space with momentum carried by vector excitations.
For example, in \cite{Mathur:2011gz, Mathur:2012tj, Lunin:2012gp}, geometries with momentum waves located at the boundary of AdS$_3$ were coupled to flat space by precisely the shift \eqref{eq:AsyFlat}.
Similarly, the harmonic functions of the singular  NS5-F1-P solution with D0-D4 dipole charges \cite{Bena:2022sge, Ceplak:2022wri} possess such constant factors, when viewed as solutions in flat space. 
However, in these two examples the U(1) gauge field was pure gauge, so in the near-horizon limit the bulk geometries become trivial: Since momentum charge carriers are localised in the transition between the near-horizon AdS$_3$ and asymptotically flat regions,  the geometry becomes deformed only when coupled to flat space. 

In vector superstrata the non-trivial microstructure is located at the bottom of the AdS$_2$ throat, so the coupling to flat space is not immediate.
While leave this interesting avenue for future work, we would like to point out that adding one can freely add a constant term to $Z_1$ only
\begin{align}
\label{eq:LinDil}
    Z_{1} \xrightarrow[]{} 1+ Z_{1}\,, \qquad  Z_{2}~~\text{unchanged}\,,
\end{align}
without spoiling the BPS equations. 
This means that in an appropriate frame one can extend the solutions beyond  AdS and into the linear-dilaton region.

\subsubsection{Conserved charges}

In this subsection we read off the asymptotic charges of the geometry and compare them to the values predicted from the analysis of the CFT states \eqref{eq:FullRCharges}.
We use the methods adapted for asymptotically flat solutions --  We assume that the addition of constant factors in the harmonic function \eqref{eq:AsyFlat} does not produce terms that contribute to the conserved charges.%
\footnote{This can happen for example if the additional terms that are necessary to solve the BPS equations all contribute at subleading order at infinity  \cite{Bena:2022sge} or are  phase dependent and thus averaging to zero \cite{Bena:2017xbt}.}
In other words, we assume that all interesting microstructure is located near the center of the geometry, far away from the transition  between the near-horizon and flat space regions.
Furthermore, if the geometries correspond states in the dual CFT then we expect that their conserved charges, which are matched to the  CFT eigenvalues, should be completely contained in the near horizon region.

We first compare the regularity condition \eqref{eq:Regularity} to the total winding constraint on CFT state \eqref{eq:StrandBudget}, which suggests the identification 
\begin{align}
	\label{eq:Trans1}
	\frac{N_a}{N} = \frac{ R_y^2\, a^2}{Q_1\,Q_5}\,,\qquad \frac{N_b}{N} =\frac{x_{k,m,n}}{2\,k}\,\frac{b^2\,R_y^2}{ Q_1\,Q_5}\,.
\end{align}
On the gravity side, the charges are extracted by analysing the asymptotic behaviour of the metric.
The fall-off of the one-forms $\omega$ and $\beta$ contains the information about the angular momenta \cite{Myers:1986un, Giusto:2013bda}
\begin{align}
    \beta_{\phi} + \beta_{\psi}+ \omega_{\phi} + \omega_{\psi}\xrightarrow[r\to \infty]{}\sqrt{2}\,\frac{J- \Jb\,\cos2\theta}{r^2}\,.
\end{align}
This allows us to read-off these charges from the geometries, and we find
\begin{align}
\label{eq:SugraJJb}
    J = \frac{a^2\, R_y}2 + \frac{m\,x_{k,m,n}}{k}\,\frac{b^2\,R_y}{2}\,,\qquad \bar J = \frac{a^2\, R_y}{2}\,.
\end{align}
The momentum charge along the $S^1_y$-circle is obtained by expanding
\begin{align}
		\cft \xrightarrow{r\to \infty} - \frac{2\,Q_P}{r^2}\,.
\end{align}
We can extract $Q_P$ using 
\begin{align}
	F_{2k,2m,2n}\xrightarrow{r\to \infty} \left[- \frac{(k-m)!\,(m-1)!n!}{4\,k\,(k+n)!}\right]\,\frac{1}{r^2} + \cO\left(r^{-3}\right)\,,
\end{align}
from which it follows that
\begin{align}
	\cft^{(k,m,n)} \xrightarrow{r\to \infty} -\frac{b^2}{r^2}\, \frac{x_{k,m,n}}{k}\,(m+n)\,,
\end{align}
and thus
\begin{align}
\label{eq:SugraMom}
    Q_P = \frac{x_{k,m,n}}{2k}\,b^2\,(m+n)\,.
\end{align}
Assume that we are working in a ten-dimensional frame in which the global charges are D1, D5, and P, for example in the frame presented in Section~\ref{ssec:D1D5P}.
In this case, the supergravity and the quantised charges in the D1-D5 CFT 
 then the supergravity charges related by
\begin{align}
 j =  \frac{N\, R_y}{Q_1\,Q_5}\, J\,,\qquad   \jb =  \frac{N\, R_y}{Q_1\,Q_5}\, \Jb\,,\qquad n_P =  \frac{N\, R_y^2}{Q_1\, Q_5}\, Q_P\,,
\end{align}
which, in addition to \eqref{eq:Trans1}, gives
\begin{align}
    j = \frac{N_a}{2} + N_b\, m\,, \qquad \jb = \frac{N_a}{2}\,,\qquad n_P=  N_b \left(m+n\right)\,.
\end{align}
These values are in precise agreement with \eqref{eq:FullRCharges} and \eqref{eq:MomChargeCFT}, retroactively justifying the use of the techniques in this section. 
This matching provides a non-trivial check that the geometries build in the preceding sections are really to the (coherent superposition of) states \eqref{eq:FullStateR}.
However, a more detailed  detailed holographic analysis is needed to fully confirm this match.

\section{Singular corner of parameter space}
\label{sec:ato0}

We want to analyse the behaviour of $(1,0,n)$ vector superstrata in the limit where the dual CFT state does not contain any $\kett{++}^{\rm R}_1$.
In the CFT, this is the limit $N_a \to 0$ and the result is a well-defined state with vanishing $SU(2)$ charges $j = \jb = 0$.
Through the identification \eqref{eq:Trans1}, this corner of parameter space corresponds to sending $a \to 0$, which is the limit in which all string sources are collapsed to a single point.
The resulting geometries would have $J = \Jb =0$ and the $SO(4)$ symmetry of $\IR^4$ is recovered. 

\begin{figure}[t]
	\includegraphics[width=\textwidth]{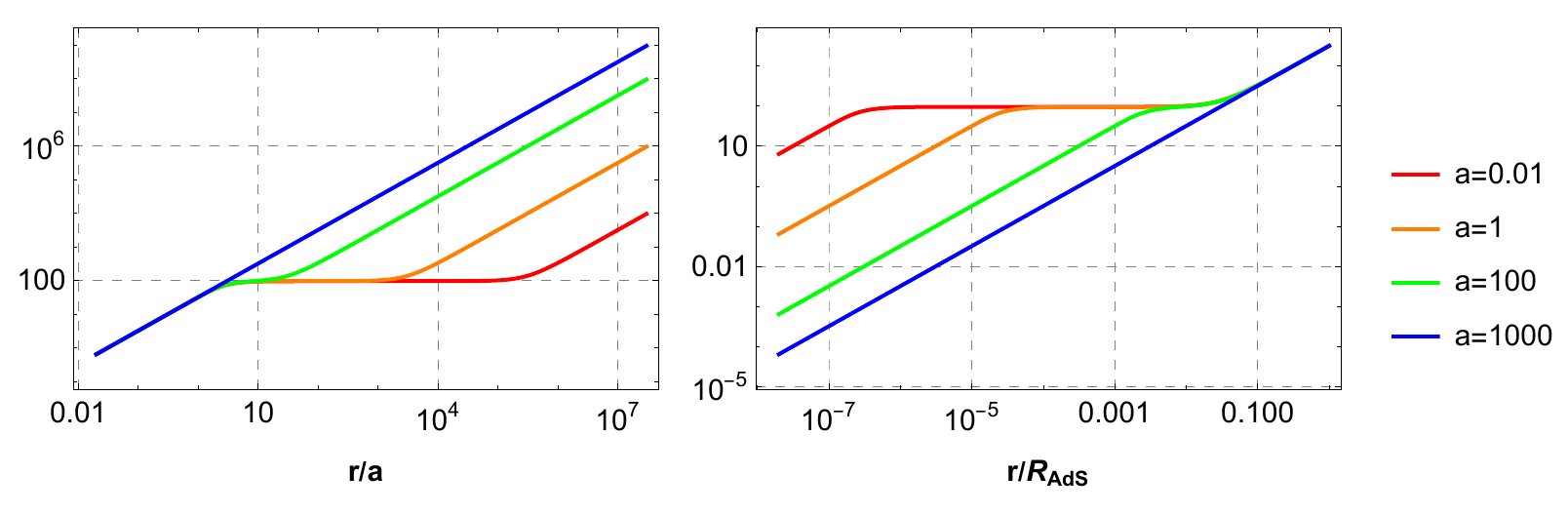}
	\label{fig:109a}
  \vspace{-0.5cm}
	\caption{The log-log plots of $\sqrt{f_y}$, for  $(k,m,n) = (1,0,9)$  superstrata as a function of the radial coordinate.
	We take $\theta = \tfrac{\pi}4$ and $Q_1 = Q_5 = 10^{10}$, $R_y = 10^7$, and plot for different values of $a$. 
	At the maximal value $a=1000$, the regularity condition imposes $b=0$ and we retrieve the unperturbed supertube without an AdS$_2$ throat (depicted in blue). 
	On the left we see that the throat region, characterised by a constant size of the $y$-circle, always develops around $r\sim a$.
	When measured in units of the radius of AdS, the length of the throat increases with decreasing $a$. At $a =0$ the length of the throat is infinite and the geometry develops an extremal horizon.}
\end{figure}

We are interested in the bulk dual of states \eqref{eq:FullStateR} in the regime of parameters 
\begin{align}
	\label{eq:Limit}
    N_a \ll N_b\,, \qquad N~\text{fixed}.
\end{align}
For that purpose, let us define 
\begin{align}
    x^2 \equiv \frac{N_a}{N_b} =2\,n\,(n+1)\, \frac{a^2}{b^2}\,,
\end{align}
where we used \eqref{eq:Trans1} and expand the metric \eqref{eq:MetAns1} in small $x$. In Figure~\ref{fig:109a}, we show that as $a$, and this $x$, is decreased, the length of the region where the $S^1_y$ is stabilised increases. 
The metric to leading order is given by
\begin{align}
\label{eq:BTZbh}
    ds^2 = - \frac{\rt^2f(\rt)^2 }{\sqrt{Q_1\,Q_5}}\, dt^2 + \frac{\sqrt{Q_1\,Q_5} }{\rt^2 f(\rt)^2}\, d\rt^2 + \frac{\rt^2}{\sqrt{Q_1\,Q_5}}\left(dy + \frac{Q_P}{\rt^2}\,dt\right)^2+ \sqrt{Q_1\,Q_5}\, d\Omega_3^2\,,
\end{align}
where $d\Omega_3^2$ is the metric on a unit 3-sphere,
$Q_P = Q_P^{(1,0,n)} = b^2/2(n+1)$, and we defined \cite{Hyun:1997jv} 
\begin{align}
    r^2 \equiv \rt^2 - Q_P\,, \qquad f(\rt) \equiv 1- \frac{Q_P}{\rt^2}\,.
\end{align}
This is exactly the metric of an extremal BTZ black hole with mass 
\begin{align}
    M= \frac{2 Q_P}{\sqrt{Q_1\,Q_5}} = \frac{b^2}{(n+1)\,\sqrt{Q_1\,Q_5}}\,,
\end{align}
multiplied by a three-sphere with constant radius. 
In addition, the dilaton field becomes a constant,  the vector field vanishes, and the three-form gauge field strength reduces to the sum of volume forms.%
\footnote{In the supersymmetric ansatz this solution corresponds to $Z_1 = \frac{Q_1}{r^2}$, $Z_2= \frac{Q_5}{r^2}$, $\cF = \cft = -2 \frac{Q_P}{r^2}$, with all other quantities vanishing.}
Thus, when $a$ vanishes, the initially horizonless geometry degenerates: It develops an infinitely long AdS$_2$ throat and an extremal horizon.

The appearance of an extremal horizon in this singular corner of parameter space was already observed in superstrata based on tensor fields \cite{Bena:2022sge}. 
There, it was argued that since taking $a=0$ corresponds to shrinking the supertube to a point, all excitations that are sourced along the locus of the supertube should vanish.%
\footnote{This is equivalent to saying that there is only a limited number of $SO(4)$ invariant excitations of global AdS$_3\times S^3$ in supergravity \cite{Kanitscheider:2007wq}.}
Indeed, taking $a\to0$ turned off the momentum carrying excitations, but it did not remove their effects -- the geometry still had momentum -- and this inconsistency is characterised by the horizon.
We see that the same argument holds in the case of vector superstrata: The vector fields that carry momentum are sourced along the supertube and when $a$ is set to 0, these fields vanish without the momentum charge decreasing, as can be seen explicitly in \eqref{eq:BTZbh}.

One may wonder whether this extremal horizon can be removed by the inclusion of appropriate degrees of freedom.
In fact, in \cite{Bena:2022sge}, vector fields were proposed as candidates to resolve the horizon, but unfortunately our analysis shows that vector fields alone are not sufficient. 
Recent analysis of supersymmetry projectors shows that there exist more general brane configurations that preserve 16 supersymmetries locally while being located at a single point in $\IR^4$ \cite{Bena:2022wpl, themelia}. However, these ``super-maze'' solutions necessary include fields which are not captured by the ansatz in this paper, thus providing an explanation for the appearance of the infinitely long throat in the singular limit of parameter space.

Alternatively, it may be necessary take into account the effects of modes living at the boundary of the AdS$_2$ region, which may become strongly coupled 
as the length of the AdS$_2$ throat increases and possibly prevent the formation of an infinitely long throat \cite{Lin:2022rzw, Lin:2022zxd}.
These effects should be important already at the level of supergravity and would not require any quantum corrections to the classical six-dimensional theory nor the inclusion of additional degrees of freedom.
Interestingly, the limit $a\xrightarrow[]{}0$ is also  where one expects stringy effects to become important \cite{Martinec:2020cml, Martinec:2022okx}. 
Is there an interplay or competition between stringy and quantum corrections and if so, can they resolve the horizon?
In the next section we show that in a particular ten-dimensional frame, vector superstrata are solutions of type II supergravity that are purely in the NS sector.
Therefore, they are amenable to exact worldsheet methods and may present backgrounds in which stringy and quantum effects may be analysed  in a qualitative way.

\section{From six to ten dimensions}
\label{sec:10D uplift}

In this section, we present uplifts of all supersymmetric configurations of $\cN=(1,0)$ supergravity in six-dimensions%
\footnote{Summarized in Appendix~\ref{sec:BPSeq}.} to solutions of ten-dimensional type IIA and IIB supergravity compactified on a four torus.
We present several frames which are related through S and T-dualitites. 

We follow the conventions summarized in Appendix~B of \cite{Bena:2022sge}.
Throughout this section, we use the democratic formalism to describe the ten-dimensional fields \cite{Bergshoeff:2001pv}, as it is particularly useful when working with brane systems. 
This means that in addition to the NS-NS two-form gauge field $B_2$, we allow for additional R-R gauge potentials, which are $\{C_1, C_3, C_5,C_7\}$ in type IIA and $\{C_0, C_2, C_4,C_6, C_8\}$ in type IIB theory, subject to the constraints
\begin{subequations}
	\label{eq:DemFormSD}
	\begin{align}
		&(IIA):&&  F_2 = * F_8 \,, \quad F_4 = - * F_6\,, \quad F_6 = * F_4\,, \quad F_8 = - * F_2\,,\\*
		&(IIB):&& F_1=*F_9,\quad F_3=-*F_7, \quad F_5=*F_5,\quad F_7=-*F_3,\quad F_9=*F_1\,, 
	\end{align}
\end{subequations}
where
\begin{align}
	F_{n+1} \equiv d C_{n}- H_3\wedge C_{n-2}\,,\qquad H_3 \equiv d B_2\,.
\end{align}
These constraints ensure the right number of degrees of freedom by imposing that $F_p$ and $F_{10-p}$ essentially contain the same information.

Throughout this section form-fields with a subscript (i.e. $B_2$, $F_2$) denote fields in the ten-dimensional theory. Form-fields without subscripts (i.e. $B$, $F$) denote quantities in six dimensions. The ansatz quantities, such as $Z_1$ or $\gamma_1$ denote the same quantity in both theories. 

\subsection{F1-NS5-P-(D0-D4-D2-D6)}

We begin by assuming that $T^4$ participates in the dynamics only through its volume form $\volt$.
Furthermore, we assume that the global charges in the system are given by fundamental strings and NS5-branes, where the former are smeared and the latter wrap the four-torus. 
This naturally leads us to the ansatz
\begin{subequations}
	\label{eq:Ans10to6}
	\begin{gather}
		ds_{(10)}^2 = e^{\phi}\,g_{\mu\nu}\, dx^\mu\,dx^\nu + \delta_{ab}\, dz^a\,dz^b\,,\\
		F^{(6)} \equiv F^{(2)}\wedge \volt\,, \qquad F^{(8)} \equiv F^{(4)}\wedge \volt\,,
	\end{gather}
\end{subequations}
where $g_{\mu\nu}$ is the Einstein-frame metric in six-dimensions.
The constraints of the democratic formalism then imply\footnote{The six-dimensional Hodge dual is always taken with respect to the Einstein-frame metric \eqref{eq:MetAns1}, unless explicitly indicated otherwise.}
\begin{align}
	F_2 = e^{-\phi}\, *_6F_4\,, \qquad F_4 =- e^{\phi}\, *_6F_2\,,
\end{align}
from which it follows within this ansatz, all degrees of freedom contained within the R-R gauge fields come from a single vector field, $C_1$. 
If we now identify
\begin{align}
	\label{eq:6Dto10DIdent}
	H_3 \equiv - e^{- 2\phi} *_6 G\,,\qquad F_2 \equiv F\,,\qquad \phi_{10} \equiv \phi\,,
\end{align}
or equivalently
\begin{align}
	B_2 \equiv 	\widetilde B\,,\qquad C_1 \equiv A\,,
\end{align}
then one can show (see Appendix~\ref{app:10to6}) that the equations of motion in ten dimensions reduce to those of six-dimensional $\cN=(1,0)$ supergravity coupled to a tensor and vector multiplet. 
Since the supersymmetric configurations that are summarised in Appendix~\ref{sec:BPSeq} solve the six-dimensional equations of motion, they  also solutions of the ten-dimensional theory. 
Hence we can use the above identifications to find the ansatz for fields in the ten-dimensional theory
\begin{subequations}
	\label{eq:10DAnsatz1}
	\begin{align}
		ds^2 &= - \frac{2}{Z_1}\,(d v+\beta)\Big[d u+\omega + \frac{\cft}{2}(d v+\beta)\Big]+Z_2\,d s^2_4+ d\hat s_4^2\,,\\
		e^{2\phi_{10}} & = \frac{Z_2}{Z_1}\,,\\
		B_2 &= -\frac{1}{Z_1}\, (du +\omega)\wedge (dv + \beta)+ a_1 \wedge (dv + \beta) + \gamma_2\,,\\
		C_1 & = \frac{Z_A}{Z_2}\,(dv + \beta) - \tA\,,\\
		C_3 &= \frac{1}{Z_1}\,\tA \wedge(du +\omega)\wedge (dv + \beta) + \delta_2 \wedge (dv+ \beta)+ x_3\,,\\
		C_5 &= C_1 \wedge \volt - \frac{1}{Z_1}\,x_3\wedge (du +\omega)\wedge (dv + \beta)\,,\\
		C_7 &= C_3 \wedge \volt\,,
	\end{align}
\end{subequations}
where the metric is given in the ten-dimensional string frame and $d\hat s_4^2$ is the flat metric on $T^4$.
We used the same ansatz quantities as in the six-dimensional ansatz, with the addition of a a two-form $\delta_2$ and a three-form, $x_3$, both with legs only on the base space, which satisfy
\begin{subequations}
	\label{eq:x3Eq}
	\begin{gather}
		\label{eq:AConstEq1}
		*_4\left(\cD Z_A + Z_2\, \dot \tA\right) = \cD \delta_2 - \delta_2 \wedge \dot \beta - \dot x_3 - \Theta^1 \wedge \tA\,\\
		\label{eq:ConstEq2}
		\cD x_3 + \delta_2 \wedge \cD\beta = \tA \wedge *_4 \left( \cD Z_2 + Z_2\,\dot \beta\right)\,.
	\end{gather}
\end{subequations}

A brane construction that realises this system in type IIA supergravity consists of global F1, NS5, P, and dipolar D0, D2, D4, and D6-brane charges.
In Table~\ref{tab:sumAns}, we summarise this brane content and the ansatz quantity to which each constituent is associated with.
\begin{table}[t!]
	\centering
	\begin{tabular}{|c||c|c|c|c|}
		\hline
		\thead{Ansatz\\ quantity}	& \thead{F1-NS5-P\\ (D0-D4-D2-D6)}&	 \thead{F1-NS5-P\\ (D1-D3-D3-D5)}& \thead{D5-D1-P\\ (P-D1-D5-KKM)} & \thead{NS5-F1-P\\ (P-F1-NS5-KKM)}\\
		\hline
		$Z_1$ & F1($y$) & F1($y$) &  D5($y6789$) &  NS5($y6789$) \\
		\hline
		$Z_2$ & NS5($y6789$) & NS5($y6789$) &D1($y$) &F1($y$)\\
		\hline
		$ - \cF$ & P($y$) & P($y$) &P($y$) &P($y$) \\
		\hline
		$Z_A$ & \thead{D4($6789$)\\D0} & \thead{D3($678$)\\D1($9$)}  & \thead{D1($9$)\\P($9$)}  & \thead{F1($9$)\\P($9$)} \\
		\hline
		$a_1$ ($\sim\Theta^1$)  & F1($\chi$) & F1($\chi$) & D5($\chi6789$)  &NS5($\chi6789$)  \\
		\hline
		$a_2$ ($\sim\Theta^2$)& NS5($\chi6789$) & NS5($\chi6789$) & D1($\chi$) & F1($\chi$)\\
		\hline
		$\tA$ & \thead{D2($y\chi$)\\D6($y\chi6789$)}  & \thead{D3($y\chi9$)\\D5($y\chi678$)} &\thead{D5($y\chi678$)\\KKm($9$)}&\thead{NS5($y\chi678$)\\KKm($9$)} \\
		\hline
	\end{tabular}
	\caption{
		The dictionary between the ansatz quantities and the brane content in each frame.
		In the parenthesis we denote the directions in which the brane are extended, with $6,7,8,9 \in T^4$ and $\chi$ denoting an arbitrary direction in $\IR^4$. 
		For the Kaluza-Klein monopole (KKm) we only denote the ``special'' direction it fibers (see for example \eqref{eq:NSNSSystemLin}).
		Note that in the last column, the role of $Z_1$ and $Z_2$ is interchanged compared to the first two frames.
		The pairs of brane-charges described by $Z_A$ and $\tA$ are locked by supersymmetry \cite{themelia}.}
	\label{tab:sumAns}
\end{table}

\subsection{F1-NS5-P-(D3-D1-D5-D3)}

Starting from \eqref{eq:10DAnsatz1}, we can perform several S-dualities and T-dualities to obtain supersymmetric solutions in different frames.
If we dualise along a direction of the $T^4$, which we take to be $z_9$, then we land in a type IIB theory with the same global charges

\begin{footnotesize}
	\begin{equation}
		\label{eq:Chain1}
		\begin{pmatrix}
			\begin{array}{c}
				\text{F1}(y)\\
				\text{NS5$(y6789)$}\\
				P(y)\\
				\hline
				\text{D4$(6789)$}\\
				\text{D0}\\
				\text{D6$(y\chi 6789)$}\\
				\text{D2$(y\chi)$}
			\end{array}
		\end{pmatrix}_{\rm IIA}\hspace*{-0.4cm}\,\,
		\xleftrightarrow[]{~\text{T}(9)~}~
		\begin{pmatrix}
			\begin{array}{c}
				\text{F1}(y)\\
				\text{NS5$(y6789)$}\\
				P(y)\\
				\hline
				\text{D3$(678)$}\\
				\text{D1$(9)$}\\
				\text{D5$(y\chi 678)$}\\
				\text{D3$(y\chi 9)$}	
			\end{array}
		\end{pmatrix}_{\rm IIB}	\,,
	\end{equation}
\end{footnotesize}%
where in the brackets we denote the directions in which the charges are extended, the top three entries correspond to global charges, while those below the horizontal line are dipole charges.
Performing this T-duality on the ansatz 
\begin{subequations}
	\label{eq:Tdual1}
	\begin{align}
		ds^2 &= - \frac{2}{Z_1}\,(d v+\beta)\Big[d u+\omega + \frac{1}{2}\,\left(\cF + \frac{Z_A^2}{Z_2}\right)(d v+\beta)\Big]+Z_2\,d s^2_4+ d\hat s_4^2\,,\\
		e^{2\phi_{10}} & = \frac{Z_2}{Z_1}\,,\\
		B_2 &= -\frac{1}{Z_1}\, (du +\omega)\wedge (dv + \beta)+ a_1 \wedge (dv + \beta) + \gamma_2\,,\\
		C_0 &= 0\,,\\
		C_2 & = \left(\frac{Z_A}{Z_2}\,(dv + \beta) - \tA\right)\wedge dz^9\,,\\
		C_4 &= \left(\frac{1}{Z_1}\,\tA \wedge(du +\omega)\wedge (dv + \beta) + \delta_2 \wedge (dv+ \beta)+ x_3\right)\wedge dz^9\nonumber\\&\quad + \left(\frac{Z_A}{Z_2}\,(dv + \beta) - \tA\right)\wedge dz^6\wedge dz^7\wedge dz^8 \,,\\
		C_6 &=\left(\frac{1}{Z_1}\,\tA \wedge(du +\omega)\wedge (dv + \beta) + \delta_2 \wedge (dv+ \beta)+ x_3\right)\wedge dz^6\wedge dz^7\wedge dz^8  \nonumber\\&\quad- \frac{1}{Z_1}\,x_3\wedge (du +\omega)\wedge (dv + \beta)\wedge dz^9 \,,\\
		C_8 &= 0\,.
	\end{align}
\end{subequations}
In this frame the vector field degree of freedom is to linear order still contained only in the R-R sector, while fields in the NS-NS sector, including the metric,  get deformed only at non-linear order.
We note that from the worldsheet analysis this seems to be the correct frame to interpret the states \eqref{eq:states2} \cite{Martinec:2022okx}. 
Interestingly, since in this frame $Z_1$ contains the information about the NS5-brane charges, the trivial addition of the constant factor in \eqref{eq:LinDil} allows us to extend the asymptotically AdS$_3$ solution to the linear dilaton region.
Finally, let us note that due to T-duality along a single direction on the four-torus, we have  singled out of a specific direction and explicitly broken $T^4$ invariance.
But while the forms and metric might have legs along the $z^9$ direction, their components cannot depend on this coordinate. 
This is a common feature of all subsequent frames.

\subsection{D1-D5-P-(D1-P-KKm-D5)}
\label{ssec:D1D5P}

The two ansatze considered so far do not highlight the importance of rewriting the BPS equations in the linear formulation, as for example, the $\cft$ combination can be clearly seen in the metric.
However, consider the following chain of dualities

\begin{footnotesize}
	\begin{equation}
		\label{eq:ChainNSNS}
		\begin{pmatrix}
			\begin{array}{c}
				\text{F1}(y)\\
				\text{NS5$(y6789)$}\\
				P(y)\\
				\hline
				\text{D3$(678)$}\\
				\text{D1$(9)$}\\
				\text{D5$(y\chi 678)$}\\
				\text{D3$(y\chi 9)$}	
			\end{array}
		\end{pmatrix}_{\rm IIB}	
		\hspace*{-0.5cm}\,\,
		\xleftrightarrow[]{~\text{S}~}~
		\begin{pmatrix}
			\begin{array}{c}
				\text{D1}(y)\\
				\text{D5$(y6789)$}\\
				P(y)\\
				\hline
				\text{D3$(678)$}\\
				\text{F1$(9)$}	\\
				\text{NS5$(y\chi 678)$}\\
				\text{D3$(y\chi9)$}
			\end{array}
		\end{pmatrix}_{\rm IIB}\hspace*{-0.5cm}\,\,
		\xleftrightarrow[]{~\text{T}(T^4)~}~
		\begin{pmatrix}
			\begin{array}{c}
				\text{D5$(y6789)$}\\
				\text{D1}(y)\\
				P(y)\\
				\hline
				\text{D1($9$)}\\
				\text{P$(9)$}	\\
				\text{KKm$(9)$}\\
				\text{D5$(y\chi678)$}
			\end{array}
		\end{pmatrix}_{\rm IIB}\,.
	\end{equation}
\end{footnotesize}%
The ansatz in the final frame reads 
\begin{subequations}
	\label{eq:Tdual9n876vinv}
	\begin{align}
		ds^2 &= - \frac{2}{\sqrt{Z_1\,Z_2}}\,(d v+\beta)\left[d u+\omega + \frac{\cft}{2}\,(d v+\beta)\right]+\sqrt{Z_1\,Z_2}\,d s^2_4\nonumber\\*
		& \quad+\sqrt{\frac{Z_2}{Z_1}}\, \left[dz^6\,dz^6+dz^7\,dz^7+dz^8\,dz^8+\left(dz^9-\frac{Z_A}{Z_2}\,(dv + \beta) + \tA\right)^2\right]\,,\\
		e^{2\phi_{10}} & = \frac{Z_2}{Z_1}\,,\\
		B_2 &=0\,,\\
		C_0 &= 0\,,\\
		C_2 &=  \frac{1}{Z_2}(du +\omega)\wedge (dv + \beta)- \at_2 \wedge (dv + \beta) - \gamma_1 - \left(\frac{Z_A}{Z_2}\,(dv + \beta) - \tA\right)\wedge dz^9 \,,\\
		C_4 &= 0\,,\\
		C_6 & = \left[\frac{1}{Z_1}\, (du +\omega)\wedge (dv + \beta)- a_1 \wedge (dv + \beta) - \gamma_2\right]\wedge\volt\nonumber\\*
		& \quad -\left(\frac{1}{Z_1}\,\tA\wedge\, (du +\omega)\wedge (dv + \beta)+(dv + \beta)\wedge\delta_2 + x_3\right)\wedge dz^6\wedge dz^7\wedge dz^8\,,\\
		C_8 &= 0\,.
	\end{align}
\end{subequations}
In this frame, the six-dimensional vector field arises as a component of the metric and a component of the two-form R-R field.
The ansatz can be written more compactly written using the field redefinitions \eqref{eq:ReDef}
\begin{subequations}
	\label{eq:Tdual9n876}
	\begin{align}
		ds^2 &= - \frac{2}{\sqrt{Z_1\,Z_2}}\,(d v+\beta)\left[d u+\omega + Z_A \big(dz^9+ \tA\big)+ \frac{\cF}{2}\,(d v+\beta)\right]+\sqrt{Z_1\,Z_2}\,d s^2_4\nonumber\\*
		& \quad+\sqrt{\frac{Z_2}{Z_1}}\, \left[dz^6\,dz^6+dz^7\,dz^7+dz^8\,dz^8+\big(dz^9+ \tA\big)^2\right]\,,\\
		e^{2\phi_{10}} & = \frac{Z_2}{Z_1}\,,\\
		B_2 &=0\,,\\
		C_0 &= 0\,,\\
		C_2 &=  \frac{1}{Z_2}\left[du +\omega+ Z_A \big(dz^9+ \tA\big)\right]\wedge (dv + \beta)- a_2\wedge (dv + \beta) - \gamma_1+  \tA\wedge \big(dz^9+ \tA\big) \,,\\
		C_4 &= 0\,,\\
		C_6 & = \left[\frac{1}{Z_1}\, (du +\omega)\wedge (dv + \beta)- a_1 \wedge (dv + \beta) - \gamma_2\right]\wedge dz^6\wedge dz^7\wedge dz^8\wedge\left(dz^9 + \tA\right)\nonumber\\*
		& \quad -\left[(dv + \beta)\wedge\left(\delta_2 + \tA\wedge a_1\right) + x_3+\gamma_2\wedge \tA\right]\wedge dz^6\wedge dz^7\wedge dz^8\,,\\
		C_8 &= 0\,.
	\end{align}
\end{subequations}
Not only can we recognise the contributions from $B_2$ and $C_3$ given in \eqref{eq:10DAnsatz1}, but we can more clearly identify the contributions from different ingredients listed in the third column of table~\ref{tab:sumAns}.
In particular, $\tA$ can be recognised as the KKm fibration along the $z^9$-direction and $Z_A$ can be understood as a sort of angular momentum along this direction. 

\subsection{NS5-F1-P-(F1-P-KKm-NS5)}

A final S-duality results in an ansatz that is completely in the NS sector for the theory

\begin{footnotesize}
	\begin{align}
		\label{eq:SDual}
		\begin{pmatrix}
			\begin{array}{c}
				\text{D5$(y6789)$}\\
				\text{D1}(y)\\
				P(y)\\
				\hline
				\text{D1($9$)}\\
				\text{P$(9)$}	\\
				\text{KKm$(9)$}\\
				\text{D5$(y\chi678)$}
			\end{array}
		\end{pmatrix}_{\rm IIB}
		\hspace*{-0.4cm}\,\,
		\xleftrightarrow[]{~\text{S}~}~
		\begin{pmatrix}
			\begin{array}{c}
				\text{NS5}(y6789)\\
				\text{F1$(y)$}\\
				P(y)\\
				\hline
				\text{F1$(9)$}\\
				\text{P$(9)$}	\\
				\text{KKm$(9)$}\\
				\text{NS5$(y\chi678)$}
			\end{array}
		\end{pmatrix}_{\rm IIB}\,,%
	\end{align}
\end{footnotesize}%
with
\begin{subequations}
	\label{eq:NSNSSysGau}
	\begin{align}
		ds^2 &= - \frac{2}{Z_2}\,(d v+\beta)\left[d u+\omega + \frac{\cft}{2}(d v+\beta)\right]+Z_1\,d s^2_4\nonumber\\*
		& \quad+dz^6\,dz^6+dz^7\,dz^7+dz^8\,dz^8+\left(dz^9- \frac{Z_A}{Z_2}(dv + \beta)+ \tA\right)^2\,,\\
		e^{2\phi} & = \frac{Z_1}{Z_2}\,,\\
		B_2 &=   -\frac{1}{Z_2}(du +\omega)\wedge (dv + \beta)+ \at_2 \wedge (dv + \beta) + \gamma_1\,,\nonumber\\*
		& \quad + \left(\frac{Z_A}{Z_2}\,(dv + \beta) - \tA\right)\wedge dz^9\,,\\
		C^{(2n)} &= 0\,,
	\end{align}
\end{subequations}
or alternatively, using the linear formulation
\begin{subequations}
	\label{eq:NSNSSystemLin}
	\begin{align}
		ds^2 &= - \frac{2}{Z_2}\,(d v+\beta)\left[d u+\omega+ Z_A\big(dz^9+ \tA\big)+ \frac{\cF}{2}\,(d v+\beta)\right]+Z_1\,d s^2_4\nonumber\\*
		& \quad+dz^6\,dz^6+dz^7\,dz^7+dz^8\,dz^8+\big(dz^9+ \tA\big)^2\,,\\
		e^{2\phi_{10}} & = \frac{Z_1}{Z_2}\,,\\
		B_2 &=   -\frac{1}{Z_2}\left[du + \omega+ Z_A \big(dz^9+ \tA\big)\right]\wedge (dv + \beta)+ a_2 \wedge (dv + \beta) + \gamma_1- \tA\wedge \big(dz^9+ \tA\big),\\
		C^{(2n)} &= 0 \,.
	\end{align}
\end{subequations}
By comparing this result with the other type IIB frame with F1-NS-P global charges, \eqref{eq:Tdual1}, we note that the role of $Z_1$ and $Z_2$ is interchanged (see also Table~\ref{tab:sumAns}). 
In the first frame $Z_1$ describes the distribution of fundamental strings while in the second frame it describes NS5-branes and vice-versa for $Z_2$.
As discussed in the next section, it would be interesting to see whether we can effectively combine these two frames and determine whether the resulting six-dimensional system remains linear.

The uplifts \eqref{eq:NSNSSysGau} and \eqref{eq:NSNSSystemLin} are the main results of this section: the six-dimensional ansatz containing a vector field can be uplifted to a system with no R-R gauge fields excited, as the momentum is carried by dipolar F1-P or NS5-KKM charges.
A consequence of this uplift is that the geometries presented in Section~\ref{sec:ExpExam}, such as the $(1,0,n)$ superstrata given in \eqref{eq:10nFamily}, can be seen as non-trivial solutions in type II supergravity that are completely in the NS sector.%
\footnote{The ansatz \eqref{eq:NSNSSysGau} is written for type IIB supergravity, but performing a T-duality along any $T^3$ directions, puts us in a Type IIA frame without changing the ansatz.}
Therefore, such backgrounds are in principle amenable to exact worldsheet analysis, despite being a microstate geometry of a three-charge black hole.

\section{Discussion}
\label{sec:Discussion}

In this paper we described the systematic construction of superstrata in which momentum is carried by vector fields excitations, when viewed in six dimensions. 
While we presented only a limited number of explicit solutions, we believe that the lack of a solution for arbitrary mode numbers is merely a technical challenge, related to obtaining a closed-form solutions to the Laplace equations \eqref{eq:Z1Lap} and \eqref{eq:muEq}. 
We see no conceptual obstructions in the way of building the solutions for general $(k,m,n)$.

A technical aspect of construction is that all ansatz quantities which are excited at quadratic order in $\bt$  are independent of the phase $\hat v_{k,m,n}$, meaning that there is no need for ``coiffuring''.
This is because $\ttt^2$ and $\cft$ are self-coiffured:  They are combinations of ``physical'' ansatz quantities \eqref{eq:ReDef} which contain information about the brane degrees of freedom, as is demonstrated by the ten-dimensional uplifts. 
In fact, the combination appearing in the expression for $\cft$ can be recognised as the $S^1_y$-circle T-dual of the coiffuring condition for the system in which the first superstrata were built \cite{Bena:2015bea, Bena:2017xbt}.%
\footnote{T-dualising the tensor-field superstrata along the $y$-circle generates massive KK-modes which are not describable within supergravity.}

Since our analysis was guided by the CFT picture, we are able to propose which CFT states are dual to the new superstrata. 
The proposed match passes the simplest test of correctly reproducing the conserved charges, however more sophisticated checks are needed to confirm the correspondence. 
In particular, one needs to clarify the difference between the bulk duals of the CFT states \eqref{eq:states1} and \eqref{eq:states2} whose CFT charges are pairwise related by a shift $k \xrightarrow{} k+1$. 
The  solution generating technique used in this paper produces the same explicit solutions for two sets of states,  up to the aforementioned shift in $k$, at least at the level of six-dimensional supergravity.%
\footnote{See Appendix~\ref{app:SecondState} for details on how to match the conserved charges when considering the bulk dual of the state in \eqref{eq:states1}.}
It is important to keep in mind that in our construction we are assumed that at linear level in $\bt$ only the vector field is excited.
However, there may exist perturbations involving the vector field in which other fields are excited at well. 
This questions can be answered by determining the full spectrum of excitations around AdS$_3\times S^3$ \cite{Deger:1998nm, deBoer:1998ip}  in a theory involving vector multiplets, which may be simplest using modern techniques of exceptional field theory \cite{Eloy:2020uix, Eloy:2021fhc}.

It is possible that the bulk duals of the two sets in  of states,  \eqref{eq:states1} and \eqref{eq:states2},  are related in the six-dimensional picture, but should be uplifted to different ten-dimensional frames.
The analysis of \cite{Martinec:2022okx} suggests that in the F1-NS5-P frame, the states  \eqref{eq:states1} naturally fit within the purely NS-NS uplift, where the vector fields arise as components of the metric and the NS-NS two-form gauge field, while the states \eqref{eq:states2} should be uplifted to \eqref{eq:Tdual1}, with several R-R gauge fields excited. 
As we have shown, these two frames are related by a chain of S-T-S dualities, so vector superstrata solve the ten-dimensional equations of motion in either frame. 
However, these dualities also move us in the moduli space of the D1-D5 (F1-NS5) system, exemplified by the interchange of global F1-NS5 charges (see Table~\ref{tab:sumAns}), so the relation to the spectrum at the symmetric product orbifold point may not be straightforward.

Interestingly, one can imagine combining the momentum carriers described in the type IIB frames \eqref{eq:Tdual1} and \eqref{eq:NSNSSystemLin} and then descend to six dimensions.
If the vector fields in the two systems are independent, then the resulting six-dimensional theory should be that of supergravity coupled to a tensor and \textit{two} vector multiplets. 
Because of the interchange between the F1 and NS5 charges in the ten-dimensional frames, the vector fields in six-dimensions would couple differently to other supergravity fields, which may in particular cause the system of BPS equations to lose its upper triangular structure, which is predicted in general by the results of \cite{Cano:2018wnq}, which considered minimal supergravity in six-dimensions coupled to an arbitrary number of vector and tensor multiplets.
It  would be interesting to analyse what is the maximal number of tensor and vector multiplets that can be added to six-dimensional supergravity without losing the nice structure of BPS equations.

As already mentioned, superstrata are very special representatives of the ensemble of black-hole microstates.
However, when superstrata are sufficiently deformed one expects evolution toward more typical, stringy microstates \cite{Eperon:2016cdd, Marolf:2016nwu, Chakrabarty:2021sff, Martinec:2022okx}.
To describe such states one needs to go beyond supergravity, for example by using worldsheet techniques \cite{Martinec:2017ztd, Martinec:2018nco, Martinec:2019wzw, Martinec:2020gkv, Bufalini:2021ndn, Bufalini:2022wyp, Bufalini:2022wzu, Martinec:2022okx} which have been shown to resolve some of the singular corners of the parameter space of smooth two-charge solutions. 
These methods are most powerful when deployed in backgrounds with only NS-NS fields. 
When viewed as solutions of type IIB supergravity \cite{Giusto:2013rxa}, most superstrata involve non-trivial R-R gauge fields, which are essential as they describe the excitations that carry the momentum charge. 
Purely NS-NS superstrata have been constructed only recently \cite{Ganchev:2022exf}, however they involve non-trivial deformations of the metric, which slightly complicates their analysis. 
Since vector superstrata can be uplifted to ten-dimensional solutions that lie in the NS sector, one may be able to perform an exact a worldsheet analysis on such backgrounds. 
Vector superstrata can thus provide a window into more typical, stringy microstates.

Finally, from the perspective of ten-dimensional gravity, the geometries with vector-field excitations non-trivially involve  directions in the internal four-torus. 
Such a structure is central in the generalised  charged Weyl formalism that was recently used to construct fully analytic non-BPS microstates \cite{Bah:2020pdz, Bah:2021owp,  Heidmann:2021cms, Bah:2022pdn, Heidmann:2022zyd}. 
In those geometries, the internal directions play a vital role in ensuring the smoothness of the geometries. 
It would be interesting to see whether one can connect BPS and and non-BPS geometries in an analytic way and the ten-dimensional uplifts presented in this paper may present a first step in this direction.
%

\section*{Acknowledgements}

I would like to thank Iosif Bena,  Shaun Hampton, Anthony Houppe, Nicolas Kovensky, Emil  Martinec, Stefano Massai,  Rodolfo Russo, Masaki Shigemori, David Turton and  Nick Warner for helpful discussions.
I would also like to thank Camille Eloy and Henning Samtleben for useful correspondence. 
This work is supported by the ERC Grant 787320-QBH Structure.

\appendix


\section{Supersymmetric solutions of $\mathcal{N}=(1,0)$ supergravity with tensor and vector multiplets}
\label{sec:BPSeq}

All supersymmetric solutions of six-dimensional $\mathcal{N}=(1,0)$ minimal supergravity coupled to a tensor and a vector multiplet have been analysed in \cite{Gutowski:2003rg,Cariglia:2004kk, Cano:2018wnq}. 
We focus on the bosonic sector of the theory. 
This consists of a metric $g_{MN}$, an unconstrained two-form gauge field $B_{MN}$, a one-form vector field $A_M$, and a dilaton $\phi$. 
Solutions of the theory which preserve some supersymmetry take on a universal form.
What is more, the BPS equations determining the quantities appearing in the supersymmetric ansatz can be organised in several layers with an upper triangular form: The solutions of the previous layer act as sources for linear differential equations of the next layer  \cite{Cano:2018wnq, Ceplak:2022wri}.

In this appendix we summarize the ansatz and the BPS equations, using two equivalent formulations. 
The first uses the gauge-invariant components of the field strength and is perhaps more natural from the six-dimensional supergravity perspective. 
The second uses the gauge-dependent components of the vector field potential. 
The advantage of this formulation is that one works with simpler quantities and that the connection to the stringy origins of the solutions are more easily seen. 

\subsection{Supersymmetric ansatz in U(1) gauge-invariant form}

\subsubsection{Field decomposition}

We begin by parametrising the metric in the standard BPS form  \cite{Gutowski:2003rg,Cariglia:2004kk}:
\begin{align}
		ds_6^2  ~&=~-\frac{2}{\sqrt{Z_1\, Z_2}}(d v+\beta)\Big[d u+\omega + \frac{\widetilde{ \mathcal{F}}}{2}(d v+\beta)\Big]~+~\sqrt{Z_1\, Z_2}\,h_{mn}\, dx^m\, dx^n\,,	\label{eq:MetAns1}
\end{align}
where the null coordinates $u$ and $v$ are defined as
\begin{align}\label{eq:DefNullCoord}
	v \equiv \frac{t + y}{\sqrt 2}\,,\qquad u \equiv \frac{t-y}{\sqrt 2}\,,
\end{align}
with $y\sim y + 2\pi\,R_y$, and the four-dimensional base space is described by the coordinates $x^m$ and metric $h_{mn}$. 
This choice of coordinates is convenient because supersymmetry imposes that components of all fields are independent of one of the null coordinates, which we take to be $u$. 
In this ansazt $Z_1$, $Z_2$, $\cft$ are scalar functions, while $\omega$ and $\beta$ are one-forms on the base space, whose components can depend both on the base space coordinates and on $v$. 
The remaining bosonic fields can be decomposed as
\begin{subequations}
	\label{eq:Fields}
	\begin{align}
		e^{2\phi} ~& =~ \frac{Z_2}{Z_1}\,,\\
		F ~&  =~  (dv+ \beta) \wedge \omega_F + \tF\,,\\
		G ~&=~ d\left[ -\frac{1}{Z_2}\, (du +\omega)\wedge (dv + \beta)\right] + \widehat G_2\,,\\
		-e^{2\phi} *_6 G ~& = ~  d\left[ -\frac{1}{Z_1}\, (du +\omega)\wedge (dv + \beta)\right] + \widehat G_1\,.
	\end{align}
\end{subequations}
where $\widehat G_{1,2}$ are given by
\begin{subequations}
	\begin{align}
		\widehat G_1 ~& \equiv~ *_4 \left(\cD Z_2 + \dot \beta Z_2 \right) + (dv+ \beta)\wedge \Theta^1\,,\\
		\widehat G_2 ~& \equiv~ *_4 \left(\cD Z_1 + \dot \beta Z_1 \right) + (dv+ \beta)\wedge\widetilde\Theta^2\,.
	\end{align}
\end{subequations}
In the above, we defined a differential operator
\begin{align}
    \cD  \equiv d_4 - \beta \wedge \pd_v\,,
\end{align}
where $d_4$ is the exterior derivative restricted to the base space. 
Furthermore, we used the Hodge dual $*_4$ on the four-dimensional base space.
Note that we follow the conventions of \cite{Gutowski:2003rg,Cariglia:2004kk, Ceplak:2022wri}, where the Hodge dual of a $p$-form in $D$-dimensions is given by
\begin{align}
	\label{eq:HDDef}
	*_D X_p  ~\equiv~ \frac{1}{p! (D-p)!}\, \epsilon_{m_1\ldots m_{D-p}, n_{D-p+1} \ldots n_D}\, X^{n_{D-p+1} \ldots n_D}\, e^{m_1}\wedge \ldots e^{m_{D-p}}\,.
\end{align}
All in all, to fully characterise these fields, we need to determine a one-form $\omega_F$ and two-forms $\tF$, $\Theta^1$, and $\widetilde{\Theta}^2$, all of which have legs only along the base space.

The field strengths are defined in terms of potentials
\begin{align}
	\label{eq:FieStrDef}
	F~=~ dA\,, \qquad G ~=~ dB + F\wedge A\,, \qquad 	-e^{2\phi} *_6 G ~=~ d\widetilde B\,,
\end{align}
By introducing a scalar $Z_A$, one-forms $\tA$, $a_1$, $\at_2$, and two-forms $\gamma_{1,2}$, one can write the potential fields as 
\begin{subequations}
	\label{eq:Potentials}
	\begin{align}
		A ~& =~ \frac{Z_A}{Z_2}\,(dv + \beta) - \tA\,,\\*
		B ~& =~ -\frac{1}{Z_2}\, (du +\omega)\wedge (dv + \beta)+ \at_2 \wedge (dv + \beta) + \gamma_1\,,\\*
		\widetilde B ~& =~ -\frac{1}{Z_1}\, (du +\omega)\wedge (dv + \beta)+ a_1 \wedge (dv + \beta) + \gamma_2\,.
	\end{align}
\end{subequations}
For this ansatz to be consistent with the expressions for the field-strengths, the quantities must satisfy
\begin{align}
	\label{eq:U1Comp}
	\omega_F =-\dot \tA + \frac{Z_A}{Z_2} \, \dot \beta - \cD \left(\frac{Z_A}{Z_2}\right)\,,\qquad 
	\tF =- \cD \tA + \frac{Z_A}{Z_2}\, \, \cD \beta\,,
\end{align}
for the vector field, while the three-form field strength decomposition implies
\begin{subequations}
	\begin{gather}
		\Theta^1 ~=~ \cD a_1 - \dot \beta\wedge a_1 + \dot \gamma_2\,,\\ 
		*_4\left(\cD Z_2 + Z_2\,\dot \beta\right) ~=~ \cD \gamma_2 - a_1 \wedge \cD\beta\,,
	\end{gather}
\end{subequations}
and
\begin{subequations}
	\label{eq:Cond2}
	\begin{gather}
		\widetilde\Theta^2 ~=~ \cD \at_2  - \dot \beta \wedge  \at_2 + \dot \gamma_1 + \tA \wedge \omega_F + \frac{Z_A}{Z_2}\, \tF  \\
		*_4\left(\cD Z_1 + Z_1\,\dot \beta\right) ~=~ \cD \gamma_1  - \at_2  \wedge \cD\beta- \tF\wedge \tA\,.
	\end{gather}
\end{subequations}

\subsubsection{BPS equations}

%
Let us begin by noting that this system contains several gauge symmetries which can be used to make convenient choices based on the system at hand \cite{Ceplak:2022wri}. %
However, we can choose to work directly with gauge invariant quantities $Z_{1,2}$, $\Theta^1$, $\widetilde \Theta^2$, $\omega_F$ and $\tF$, in which case this choice is in principle irrelevant. 
The BPS equations are organised into several layers. 

\paragraph{The zeroth layer.}
One begins by determining the base space metric and the one-form $\beta$. 
This subset contains the only non-linear equations of the whole system: It imposes that the base-space must be almost hyper-K\"ahler, with the equations determining the complex structures $J^A$, $A=1,2,3$, 
\begin{subequations}
	\label{eq:Lay0J1}
	\begin{align}
		*_4 J^A ~&=~ - J^A\,,\\*
		(J^A)^m{}_{n}(J^B)^n{}_{p}~&=~ \epsilon^{ABC}(J^C)^m{}_{p}-\delta^{AB}\delta^m_p\,,\\*
		J^A \wedge J^B ~&=~ - 2\delta^{AB}\, {\rm vol}_4\,,\\*
		d_4 J^A ~&=~ \pd_v\left(\beta \wedge J^A\right)\,,
	\end{align}
\end{subequations}
and the self-duality condition for the fibration vector $\beta$
\begin{align}
	\label{eq:Dbetaeq1}
	*_4 \cD\beta ~=~ \cD\beta\,.
\end{align}
We also define an anti self-dual two-form:
\begin{align}
	\psi ~\equiv~  \frac{1}{8} \,\epsilon_{ABC}\, (J^A)^{mn}\, (\dot J^B)_{mn} \,J^C\,,
\end{align}
which appears in the subsequent layers. 

Taking the base-space metric and one-form $\beta$ to be $v$-independent simplifies the equations: The base space becomes hyper-K\"ahler, the equation for $\beta$ is now linear, and $\psi \equiv 0$.
This is the sector in which we will look for solutions.

\paragraph{The first layer.}
%
The equations in this layer determine the pair $(Z_2, \Theta^1)$
\begin{subequations}
	\label{eq:Lay1}
	\begin{gather}
		*_4 \Theta^1   ~=~  \Theta^1  -  2 \,  Z_2\,  \psi\,,\\
		\cD *_4 \left[ \cD Z_2 + Z_2 \,\dot \beta\right] + \Theta^1 \wedge \cD \beta ~=~ 0\,,\\
		d_4 \Theta^1 ~=~ \pd_v \left[ \beta \wedge \Theta^1 +  *_4 \left(\cD Z_2 + Z_2\, \dot \beta\right)\right]\,.
	\end{gather}
\end{subequations}
This layer is exactly the same as in the systems coupled to only tensor multiplets \cite{Bena:2011dd, Giusto:2013rxa}.

\paragraph{The second layer.}
This layer determines the components of the vector-field strength $\tF$ and $\omega_F$
\begin{subequations}
	\label{eq:Lay2}
	\begin{gather}
		*_4 \tF~=~ \tF\,,\\
		2\, \cD Z_2 \wedge *_4\omega_F \,+\, Z_2\, \cD *_4 \omega_F~=~ - \tF \wedge \Theta^1\,.
	\end{gather}
\end{subequations}
The first equation imposes the two-form to be self-dual.
The second equation crucially contains information about the solutions of the first layer, setting a definite order in which the layers need to be solved. 

\paragraph{The third layer.}
The next subset of equations determines $Z_1$ and $\widetilde \Theta^2$:
\begin{subequations}
	\label{eq:Lay3}
	\begin{gather}
		*_4 \widetilde\Theta^2   ~=~  \widetilde\Theta^2  -  2 \,  Z_1\,  \psi\,,\\
		\cD *_4 \left[ \cD Z_1 + Z_1 \,\dot \beta\right] + \widetilde\Theta^2 \wedge \cD \beta ~=~  \tF\wedge \tF \,,\\
		d_4 \widetilde\Theta^2 ~=~ \pd_v \left[ \beta \wedge \widetilde\Theta^2 +  *_4 \left(\cD Z_1 + Z_1\, \dot \beta\right)\right]- 2\, \omega_F \wedge \tF\,.
	\end{gather}
\end{subequations}
Note that the solutions of the previous layer appear as quadratic sources, and therefore the backreaction of any vector field perturbation will contain non-trivial $Z_1$ and $\widetilde \Theta^2$ fields. 

\paragraph{The fourth layer.}
The last two equations determine $\cft$ and $\omega$, which contain the information about the momentum along the $y$-direction and angular momentum in the $\IR^4$ respectively.
One finds
\begin{align}
	\label{eq:Lay4Eq1}
	\cD \omega + *_4 \cD \omega + \widetilde\cF \,\cD \beta~=~   Z_1\, \Theta^1 + Z_2\,\widetilde\Theta^2 - 2 Z_1\,Z_2\,\psi \,,
\end{align}
and 
\begin{align}
	\label{eq:Lay4Eq2}
	&*_4 \cD *_4  \widetilde L+ 2 \,\dot \beta^m \,\widetilde  L_m  \nonumber\\*
	&= - \frac12 *_4 \left( \Theta^1 - Z_2 \, \psi\right)\wedge \left(\widetilde\Theta^2 - Z_1 \, \psi\right) +  \ddot Z_1 \, Z_2 + \dot Z_1\, \dot Z_2 + Z_1 \, \ddot Z_2  + Z_2\,\omega_F^2  \nonumber \\*
	&\quad +\frac12 Z_1 \,Z_2 *_4  \psi\wedge  \psi +  *_4  \psi \wedge \cD \omega- \frac14 Z_1\, Z_2\, \dot h^{mn}\,\dot  h_{mn} + \frac12 \pd_v \left[Z_1 \, Z_2 \,h^{mn}\dot h_{mn} \right]\,,
\end{align}
where
\begin{align}
	\widetilde	L ~\equiv ~\dot \omega + \frac{\widetilde\cF}{2} \,\dot \beta- \frac12\, \cD\widetilde \cF\,,
\end{align}
and 
\begin{align}
	\omega_F^2~ =~ \left(\omega_F\right)_m\,\left(\omega_F\right)_n\, h^{mn}\,.
\end{align}
We observe that the only explicit contribution of the additional U(1) degrees of freedom is in the $\omega_F^2$ term in the second equation.
In \eqref{eq:Lay4Eq1} the vector field contributions appear only implicitly through $Z_1$, $\widetilde \Theta^2$, and $\cft$.

\subsection{BPS equations linear form}
\label{ssec:BPSlin}

\subsubsection{Field redefinitions}

In some cases it may be advantageous to work with the quantities appearing in the decomposition of the vector-field potential, $Z_A$ and $\tA$, rather than their field-strength combinations. 
Most importantly, as shown in Section~\ref{sec:10D uplift}, using these quantities it is easier to analyse the contributions from different string and brane sources. 
Begin by defining
\begin{subequations}
\label{eq:ReDef}
    \begin{align}
        \cft &\equiv \cF + \frac{Z_A^2}{Z_2}\,,\\
        \widetilde \Theta^2 & \equiv \Theta^2+ \frac{Z_A^2}{Z_2^2}\, \cD\beta -2\, \frac{Z_A}{Z_2}\, \cD \tA\,,
    \end{align}
\end{subequations}
which also implies
\begin{align}
    \at_2 = a_2 - \frac{Z_A}{Z_2}\,\tA\,.
\end{align}
As a consequence of this shift, one now finds that 
\begin{subequations}
	\begin{gather}
		\Theta^2 ~=~ \cD a_2 - \dot \beta \wedge a_2+ \dot \gamma_1 + \dot \tA \wedge \tA\,,\\
		*_4\left(\cD Z_1 + Z_1\,\dot \beta\right) ~=~ \cD \gamma_1+ \cD \tA \wedge \tA  - a_2 \wedge \cD\beta\,.
	\end{gather}
\end{subequations}
The rest of the ansatz changes trivially with the redefinition \eqref{eq:ReDef} (see Section~5 of \cite{Ceplak:2022wri}).
However, it is important to note that now more fields are gauge dependent: To ensure that $\cft$ and $\widetilde \Theta^2$ are invariant under the U(1) symmetry,  $\cF$ and $\Theta^2$ must compensate for the gauge-dependence of the additional terms in \eqref{eq:ReDef}, making them gauge dependent.  
%

\subsubsection{The BPS equations}

Because \eqref{eq:ReDef} involves only quantities appearing in the last three layers, the BPS equations in the zeroth and first layer remain unchanged. 
The second layer equations are now
\begin{subequations}
	\label{eq:Lay2Lin}
	\begin{gather}
		*_4 \cD \tA   ~=~ \cD \tA\,,\\*
		*_4 \cD Z_A \wedge \dot \beta \,+\,2\, \cD Z_2 \wedge *_4 \dot \tA \,+\, Z_2\,  \cD * \dot\tA \,+\, \cD *_4 \cD Z_A~=~  -\cD \tA \wedge \Theta^1\,.
	\end{gather}
\end{subequations}
They impose that the  covariant derivative of $\tA$ is self dual and determine  $Z_A$ through a generalised  Laplace equation on the base space \cite{Ceplak:2022wri}.

The third layer equations, determining $Z_1$ and $\Theta^2$ are also slightly modified
\begin{subequations}
	\label{eq:Lay3Lin}
	\begin{gather}
		*_4  \Theta^2   ~=~  \Theta^2  -  2 \,  Z_1\,  \psi\,,\\
		\cD *_4 \left[ \cD Z_1 + Z_1 \,\dot \beta\right] ~+~ \Theta^2 \wedge \cD \beta ~=~ \cD \tA \wedge \cD \tA\,,\\
		d_4 \Theta^2 ~= ~\pd_v \left[ \beta \wedge\Theta^2 +  *_4 \left(\cD Z_1 + Z_1\, \dot \beta\right)\right]~-~ 2 \dot \tA  \wedge \cD \tA \,.
	\end{gather}
\end{subequations}

Finally, the last two equations are given by
\begin{align}
	\label{eq:SumLay2eq11}
	\cD \omega + *_4 \cD \omega +  \cF \,\cD\beta ~=~  Z_1\, \Theta^1 + Z_2\, \Theta^2- 2\, Z_A\, \cD\tA - 2\, Z_1\,Z_2\,\psi \,,
\end{align}
and 
\begin{align}
	\label{eq:SumLay2eq2}
	&*_4 \cD *_4   L+ 2 \,\dot \beta^m \,  L_m \nonumber\\
	&=- \frac12 *_4 \left( \Theta^1 - Z_2 \, \psi\right)\wedge \left(\Theta^2 - Z_1 \, \psi\right) +  \ddot Z_1 \, Z_2 + \dot Z_1\, \dot Z_2 + Z_1 \, \ddot Z_2 \nonumber \\
	&\quad +\frac12 Z_1 \,Z_2 *_4  \psi\wedge  \psi +  *_4  \psi \wedge \cD \omega- \frac14 Z_1\, Z_2\, \dot h^{mn}\,\dot  h_{mn} + \frac12 \pd_v \left[Z_1 \, Z_2 \,h^{mn}\dot h_{mn} \right]\nonumber\\*
	& \quad + Z_2 \,\dot \tA^2 +  \dot\tA_m \left[\cD Z_A\right]^m\,,
\end{align}
where the one-form $L$ is now
\begin{align}
	L \equiv  \dot \omega + \frac{\cF}{2} \,\dot \beta- \frac12\, \cD  \cF+ Z_A\, \dot\tA\,.
\end{align}
and the indices are raised and lowered with respect to the four-dimensional metric $h_{mn}$.

It is important to observe that even though the shifts \eqref{eq:ReDef} are highly non-trivial and mix quantities from several different layers, the resulting equations not only retain their layered upper-triangular structure, but also remain linear differential equations, with the various shifts cancelling out non-linear contributions.

\section{Matching the ten-dimensional and six-dimensional descriptions}
\label{app:10to6}

In this Appendix we present the details of how classical solutions of ten-dimensional type IIA supergravity can be reduced to solutions of six-dimensional supergravity coupled to a tensor and vector multiplet \cite{Cariglia:2004kk}.
This shows that the geometries constructed in the main part of the text are solutions to ten-dimensional equations of motion.

We work in the so called democratic formalism \cite{Bergshoeff:2001pv} and  focus only on  the bosonic sector of the theory, which in type IIA supergravity is described by the  pseudo-action\cite{Bergshoeff:2001pv, Lust:2008zd}
\begin{align}
	\label{eq:APseudoActIIA}
	S_P^{(10)} &= \frac1{2 \kappa_{10}^2}\int d^{10}x\, \sqrt{-G}\, \Biggr\{ e^{-2\phi}\left[ R\left(G\right) + 4\,\left(\pd\phi\right)^2 - \frac12\,H_3\cdot H_3\right] - \frac14\sum_{i=1}^4\,F_{2n}\cdot F_{2n}\Biggr\}\,,
\end{align}
where $G_{MN}$ denotes the ten-dimensional metric in the string frame and we used 
\begin{align}
	X_p \cdot Y_p \equiv \frac{1}{p!}\, X_{M_1\,\ldots M_p}\, Y^{M_1\,\ldots M_p}\,.
\end{align}
In addition to the metric and the dilaton, we also have a three-form NS-NS field strength $H_3$ and even-numbered R-R field-strengths $F_{2n}$, which are subject to the self-duality conditions
\begin{align}
		\label{eq:ADemFormSD}
		F_2 = * F_8 \,, \quad F_4 = - * F_6\,, \quad F_6 = * F_4\,, \quad F_8 = - * F_2\,,
\end{align}
These are related to their corresponding gauge potentials as 
\begin{align}
	\label{eq:AFielStengthDem}
	H_3 \equiv d B_2\,, \qquad F_{n} \equiv d C_{n-2}- H_3\wedge C_{n-3}\,,
\end{align}
from which we get modified Bianchi identities
\begin{align}
	\label{eq:Bianchies10D}
	dH_3 = 0\,, \qquad  dF_{n}= H_3\wedge F_{n-2}\,.
\end{align}

We begin by assuming that $T^4$ participates in the dynamics only through its volume form $\volt$.
Furthermore, we assume that the global charges in the system are given by fundamental strings and NS5-branes, where the former are smeared and the latter wrap the four-torus. 
This naturally leads us to the ansatz
\begin{subequations}
	\label{eq:AAns10to6}
	\begin{align}
		ds_{(10)}^2 &= e^{\phi}\,g_{\mu\nu}\, dx^\mu\,dx^\nu + \delta_{ab}\, dz^a\,dz^b\,,\\
		F^{(6)} &\equiv F^{(2)}\wedge \volt\,, \qquad F^{(8)} \equiv F^{(4)}\wedge \volt\,,
	\end{align}
\end{subequations}
while leaving all other fields unchanged. 
The metric $g_{\mu\nu}$ is the Einstein-frame metric in six-dimensions.
Inserting this into \eqref{eq:APseudoActIIA} reduces the pseudo-action to
\begin{align}
	\label{eq:APseudoAct6DEF}
	S_P^{(6)} &= \frac1{2 \kappa_{6}^2}\int d^{6}x\, \sqrt{-g}\, \Biggr[ R\left(g\right) - \,\left(\pd\phi\right)^2 - \frac12\,e^{-2\phi}\,H_4\cdot H_3 - \frac12\left(e^{\phi}\,F_2\cdot F_2+ e^{-\phi}\,F_4\cdot F_4\right)\Biggr]\,.
\end{align}
where $\kappa^2_{(6)}\equiv \kappa^2_{(10)}/{\rm Vol}(T^4)$.
The equations of motion, resulting from varying this pseudo-action are
\begin{subequations}
	\label{eq:AEOMn1}
	\begin{align}
		R_{\mu\nu} &= \pd_\mu\phi\,\pd_\nu\phi + \frac14 e^{-2\phi}\left(H_{\mu\alpha\beta}H_{\nu}{}^{\alpha\beta}- g_{\mu\nu}\,H_3\cdot H_3\right)+ \frac12 e^\phi\,\left(F_{\mu \alpha}\,F_{\nu}{}^{\alpha} - \frac14\,g_{\mu\nu}\,F_2\cdot F_2\right)\nonumber\\*
		& \quad + \frac12\,e^{-\phi}\left(\frac1{3!}F_{\mu\alpha\beta\gamma}F_{\nu}{}^{\alpha\beta\gamma}- \frac34\, g_{\mu\nu}\,F_4\cdot F_4\right)\,,\\
		\nabla^2\phi &= - \frac12 \,e^{-2\phi}\, H_3\cdot H_3 + \frac14 \,e^{\phi}\, F_2 \cdot F_2 - \frac14 \,e^{-\phi}\, F_4 \cdot F_4\,,
	\end{align}
\end{subequations}
for the metric and the dilaton and
\begin{subequations}
	\label{eq:AFromEqn1}
	\begin{gather}
		d\left(e^{-2\phi}*_6H_3\right) =- e^{-\phi}*_6F_4 \wedge F_2\,,\qquad d H_3 = 0\,,\\
		d\left(e^\phi*_6F_2\right) = -  e^{-\phi}*_6F_4\wedge H_3\,,\qquad d\left(e^{-\phi}*_6F_4\right) =0\,,
	\end{gather}
\end{subequations}
for the gauge fields.%
\footnote{The six-dimensional Hodge dual is always taken with respect to the metric $g_{\mu\nu}$ in the Einstein frame.}

The self-duality conditions \eqref{eq:ADemFormSD} imply that the R-R gauge fields are related via
\begin{align}
	F_2 = e^{-\phi}\, *_6F_4\,, \qquad F_4 =- e^{\phi}\, *_6F_2\,,
\end{align}
which allwos us to eliminate $F_4$ from the equations of motion.
Furthermore, we define a new three-form field strength $G$ as%
\footnote{Following \eqref{eq:FieStrDef}, this identification also implies $\widetilde B \equiv B_2$.}
\begin{align}
	\label{eq:AIndenG}
	 G \equiv - e^{-2\phi}*_6 H_3\,,\qquad H_3 \equiv - e^{2\phi} *_6G\,,
\end{align}
and identify
\begin{align}
	\label{eq:AIdentF}
		F \equiv F_2\,.
\end{align}
In this case, the equations of motion become
\begin{subequations}
	\label{eq:AEOMf}
	\begin{align}
		&R_{\mu\nu} = \pd_\mu\phi\,\pd_\nu\phi + \frac14 e^{2\phi}\left(G_{\mu\alpha\beta}G_{\nu}{}^{\alpha\beta}- g_{\mu\nu}\,G\cdot G\right)+ \frac14 e^\phi\,\left(4F_{\mu \alpha}\,F_{\nu}{}^{\alpha} - \,g_{\mu\nu}\,F\cdot F\right)\,,\\
		&\nabla^2\phi =  \frac12 \,e^{2\phi}\, G\cdot G + \frac12 \,e^{\phi}\, F \cdot F\,,\\
		&dG  = F \wedge F\,,\qquad d \left( e^{2\phi} *_6G\right) = 0\,,\\
		&d\left(e^\phi*_6 F\right) = e^{2\phi} *_6G \wedge F\,,\qquad dF =0\,.
	\end{align}
\end{subequations}
These equations arise by varying the following Lagrangian density 
\begin{align}
	\label{eq:ALagrangian}
	e^{-1} \, \cL ~=~ \frac14\,R -\frac14\, \left(\pd\phi\right)^2- \frac18\, e^{2 \phi}\, G\cdot G-\frac14\, e^\phi \, F\cdot F\,,
\end{align}
where the gauge-field strengths are defined by
\begin{align}
	\label{eq:FieldStrengthsDef}
	F ~\equiv~ d A\,,\qquad G ~\equiv~ dB + F \wedge A\,.
\end{align}
This theory is the starting point of the supersymmetry analysis of \cite{Ceplak:2022wri}. 
The ansatz presented in Appendix~\ref{sec:BPSeq} describes all supersymmetric solutions of this theory. 
As such, these are also solutions of ten-dimensional  Type IIA theory of supergravity.

\section{Details of the dualities}
\label{app:STdual}

In this appendix, we present the ansatz for all the intermediate steps in the chain of dualities 

\begin{footnotesize}
	\begin{equation}
		\label{eq:ChainNSNS}
		\begin{pmatrix}
			\begin{array}{c}
				\text{F1}(y)\\
				\text{NS5$(y6789)$}\\
				P(y)\\
				\hline
				\text{D3$(678)$}\\
				\text{D1$(9)$}\\
				\text{D5$(y\chi 678)$}\\
				\text{D3$(y\chi 9)$}	
			\end{array}
		\end{pmatrix}_{\rm IIB}	
		\hspace*{-0.5cm}\,\,
		\xleftrightarrow[]{~\text{S}~}~
		\begin{pmatrix}
			\begin{array}{c}
				\text{D1}(y)\\
				\text{D5$(y6789)$}\\
				P(y)\\
				\hline
				\text{D3$(678)$}\\
				\text{F1$(9)$}	\\
				\text{NS5$(y\chi 678)$}\\
				\text{D3$(y\chi9)$}
			\end{array}
		\end{pmatrix}_{\rm IIB}\hspace*{-0.5cm}\,\,
		\xleftrightarrow[]{~\text{T}(T^4)~}~
		\begin{pmatrix}
			\begin{array}{c}
				\text{D5$(y6789)$}\\
				\text{D1}(y)\\
				P(y)\\
				\hline
				\text{D1($9$)}\\
				\text{P$(9)$}	\\
				\text{KKm$(9)$}\\
				\text{D5$(y\chi678)$}
			\end{array}
		\end{pmatrix}_{\rm IIB}
	\end{equation}
\end{footnotesize}%
and in particular will expand the four T-dualities along the $T^4$.

\begin{scriptsize}
\begin{equation}
		\label{eq:AT4ChainFull}
		\begin{pmatrix}
			\begin{array}{c}
				\text{D1}(y)\\
				\text{D5$(y6789)$}\\
				P(y)\\
				\hline
				\text{D3$(678)$}\\
				\text{F1$(9)$}	\\
				\text{NS5$(y\chi 678)$}\\
				\text{D3$(y\chi9)$}
			\end{array}
		\end{pmatrix}_{\rm IIB}\hspace*{-0.6cm}\,\,
		\xleftrightarrow[]{\text{T}(9)}\begin{pmatrix}
			\begin{array}{c}
				\text{D2$(y9)$}\\
				\text{D4$(y678)$}\\
				P(y)\\
				\hline
				\text{D4($6789$)}\\
				\text{P$(9)$}	\\
				\text{KKm$(9)$}\\
				\text{D2$(y\chi)$}
			\end{array}
		\end{pmatrix}_{\rm IIA}
		\hspace*{-0.6cm}\,\,
		\xleftrightarrow[]{\text{T}(8)}
		\begin{pmatrix}
			\begin{array}{c}
				\text{D3$(y89)$}\\
				\text{D3}(y67)\\
				P(y)\\
				\hline
				\text{D3($679$)}\\
				\text{P$(9)$}	\\
				\text{KKm$(9)$}\\
				\text{D3$(y\chi8)$}
			\end{array}
		\end{pmatrix}_{\rm IIB}
		\hspace*{-0.6cm}\,\,
		\xleftrightarrow[]{\text{T}(7)}
		\begin{pmatrix}
			\begin{array}{c}
				\text{D4$(y789)$}\\
				\text{D2}(y6)\\
				P(y)\\
				\hline
				\text{D2($69$)}\\
				\text{P$(9)$}	\\
				\text{KKm$(9)$}\\
				\text{D4$(y\chi78)$}
			\end{array}
		\end{pmatrix}_{\rm IIA}\hspace*{-0.6cm}\,\,
		\xleftrightarrow[]{\text{T}(6)}\begin{pmatrix}
			\begin{array}{c}
				\text{D5$(y6789)$}\\
				\text{D1}(y)\\
				P(y)\\
				\hline
				\text{D1($9$)}\\
				\text{P$(9)$}	\\
				\text{KKm$(9)$}\\
				\text{D5$(y\chi678)$}
			\end{array}
		\end{pmatrix}_{\rm IIB}
\end{equation}
\end{scriptsize}%
The ansatz in the final frame is given by \eqref{eq:Tdual9n876}.
Throughout this section, we will use the untilded ansatz quantities \eqref{eq:ReDef}.
As before, we follow the conventions of \cite{Bena:2022sge}.

\paragraph{S-dual}
The S-dual of the ansatz given in \eqref{eq:Tdual1} is given by
\begin{subequations}
	\label{eq:ASdual1}
	\begin{align}
		ds^2 &= - \frac{2}{\sqrt{Z_1\,Z_2}}\,(d v+\beta)\Big[d u+\omega + \frac{1}{2}\,\left(\cF + \frac{Z_A^2}{Z_2}\right)(d v+\beta)\Big]+\sqrt{Z_1\,Z_2}\,d s^2_4+\sqrt{\frac{Z_1}{Z_2}} d\hat s_4^2\,,\\
		e^{2\phi} & = \frac{Z_1}{Z_2}\,,\\
		B_2 &=\left(\frac{Z_A}{Z_2}\,(dv + \beta) - \tA\right)\wedge dz^9\,,\\
		C_0 &= 0\,,\\
		C_2 & =  \frac{1}{Z_1}\, (du +\omega)\wedge (dv + \beta)- a_1 \wedge (dv + \beta) - \gamma_2\,,\\
		C_4 &= \left[(dv + \beta)\wedge\left(\delta_2 - \frac{Z_A}{Z_2}\,\gamma_2+ \tA\wedge a_1\right) + x_3+\gamma_2\wedge \tA\right]\wedge dz^9\nonumber\\&\quad + \left(\frac{Z_A}{Z_2}\,(dv + \beta) - \tA\right)\wedge dz^6\wedge dz^7\wedge dz^8 \,,\\
		C_6 &= \left[ \frac{1}{Z_2}(du +\omega)\wedge (dv + \beta)- \left(a_2- \frac{Z_A}{Z_2}\,\tA\right) \wedge (dv + \beta) - \gamma_1\right]\wedge \volt\,\,,\\
		C_8 &= 0\,.
	\end{align}
\end{subequations}

\paragraph{T-dual along $z_9$}
\begin{subequations}
	\label{eq:ATdual9n}
	\begin{align}
		ds^2 &= - \frac{2}{\sqrt{Z_1\,Z_2}}\,(d v+\beta)\left[d u+\omega + Z_A \left(dz^9 + \tA\right)+ \frac{\cF}{2}\,(d v+\beta)\right]+\sqrt{Z_1\,Z_2}\,d s^2_4\nonumber\\*
		& \quad +\sqrt{\frac{Z_1}{Z_2}}\left[dz^6\,dz^6 +dz^7\,dz^7 +dz^8\,dz^8 \right]+\sqrt{\frac{Z_2}{Z_1}}\, \left(dz^9+ \tA\right)^2\,,\\
		e^{2\phi} & = \sqrt{\frac{Z_1}{Z_2}}\,,\\
		B_2 &=0\,,\\
		C_1 &= 0\,,\\
		C_3 & = \left[\frac{1}{Z_1}\, (du +\omega)\wedge (dv + \beta)- a_1 \wedge (dv + \beta) - \gamma_2\right]\wedge \left(dz^9 + \tA\right)\nonumber\\*
		& \quad +(dv + \beta)\wedge\left(\delta_2 + \tA\wedge a_1\right) + x_3+\gamma_2\wedge \tA\,,\\
		C_5 &= \left[ \frac{1}{Z_2}(du +\omega+Z_A (dz^9 + \tA))\wedge (dv + \beta)- a_2 \wedge (dv + \beta) - \gamma_1\right]\wedge dz^6\wedge dz^7\wedge dz^8\nonumber\\
		& \quad + \left(\frac{Z_A}{Z_2}\,(dv + \beta) - \tA\right)\wedge \volt\,,\\
		C_7 &= 0\,.
	\end{align}
\end{subequations}

\paragraph{T-dual along $z_8$}
\begin{subequations}
	\label{eq:ATdual9n8}
	\begin{align}
		ds^2 &= - \frac{2}{\sqrt{Z_1\,Z_2}}\,(d v+\beta)\left[d u+\omega + Z_A \left(dz^9 + \tA\right)+ \frac{\cF}{2}\,(d v+\beta)\right]+\sqrt{Z_1\,Z_2}\,d s^2_4\nonumber\\*
		& \quad +\sqrt{\frac{Z_1}{Z_2}}\left[dz^6\,dz^6 +dz^7\,dz^7  \right]+\sqrt{\frac{Z_2}{Z_1}}\, \left[dz^8\,dz^8+\left(dz^9+ \tA\right)^2\right]\,,\\
		e^{2\phi} & = 1\,,\\
		B_2 &=0\,,\\
		C_0 &= 0\,,\\
		C_2 &= 0\,,\\
		C_4 & = -\left[\frac{1}{Z_1}\, (du +\omega)\wedge (dv + \beta)- a_1 \wedge (dv + \beta) - \gamma_2\right]\wedge dz^8\wedge\left(dz^9 + \tA\right)\nonumber\\*
		& \quad +\left((dv + \beta)\wedge\left(\delta_2 + \tA\wedge a_1\right) + x_3+\gamma_2\wedge \tA\right)\wedge dz^8\nonumber\\*
		& \quad +  \left[ \frac{1}{Z_2}(du +\omega+Z_A \left(dz^9 + \tA\right))\wedge (dv + \beta)- a_2 \wedge (dv + \beta) - \gamma_1\right]\wedge dz^6\wedge dz^7\nonumber\\*
		& \quad -\left(\frac{Z_A}{Z_2}\,(dv + \beta) - \tA\right)\wedge dz^6\wedge dz^7 \wedge dz^9 \,,\\
		C_6 &= 0\,,\\
		C_8 &= 0\,.
	\end{align}
\end{subequations}
This is a particularly interesting system since the dilaton is constant throughout the geometry and the global and dipole charges are all carried by D3-branes.
As expected, only $C_4$ is excited out of all R-R gauge fields. 

\paragraph{T-dual along $z_7$}
\begin{subequations}
	\label{eq:ATdual9n87}
	\begin{align}
		ds^2 &= - \frac{2}{\sqrt{Z_1\,Z_2}}\,(d v+\beta)\left[d u+\omega + Z_A \left(dz^9 + \tA\right)+ \frac{\cF}{2}\,(d v+\beta)\right]+\sqrt{Z_1\,Z_2}\,d s^2_4\nonumber\\*
		& \quad +\sqrt{\frac{Z_1}{Z_2}}\,dz^6\,dz^6 +\sqrt{\frac{Z_2}{Z_1}}\, \left[dz^7\,dz^7+dz^8\,dz^8+\left(dz^9+ \tA\right)^2\right]\,,\\
		e^{2\phi} & = \sqrt{\frac{Z_2}{Z_1}}\,,\\
		B_2 &=0\,,\\
		C_1 &= 0\,,\\
		C_3 &= +  \left[ \frac{1}{Z_2}(du +\omega)\wedge (dv + \beta)- \left(a_2- \frac{Z_A}{Z_2}\,\tA\right) \wedge (dv + \beta) - \gamma_1\right]\wedge dz^6\,,\nonumber\\*
		& \quad +\left(\frac{Z_A}{Z_2}\,(dv + \beta) - \tA\right)\wedge dz^6\wedge dz^9 \,,\\
		C_5 & = -\left[\frac{1}{Z_1}\, (du +\omega)\wedge (dv + \beta)- a_1 \wedge (dv + \beta) - \gamma_2\right]\wedge dz^7\wedge dz^8\wedge\left(dz^9 + \tA\right)\nonumber\\*
		& \quad -\left((dv + \beta)\wedge\left(\delta_2 + \tA\wedge a_1\right) + x_3+\gamma_2\wedge \tA\right)\wedge dz^7\wedge dz^8\,,\\
		C_7 &= 0\,.
	\end{align}
\end{subequations}
Performing another T-duality along $z^6$ then gives \eqref{eq:Tdual9n876}.

\section{More examples}
\label{app:Examples}

In this appendix we present more explicit examples of geometries for different mode numbers. 
We begin by presenting in more detail the simplest two solutions, (1,0,1) and (1,1,1). 
In principle, these are already given in the main text, however, the general form is very cumbersome and so it might be useful to present the simplified versions for $n=1$.
We then state the backreacted ansatz quantities for some geometries with  $k=2$.

\paragraph{(1,0,1).} Setting $n=1$ in the solutions \eqref{eq:10nFamily} considerably simplifies the solution
\begin{subequations}
	\label{eq:101}
	\begin{align}
		 f^{(1,0,1)} &= -\frac{1}{16}\frac{a^2 + 2\,r^2}{(a^2 + r^2)^2} \,,\\
		Z_1^{(1,0,1)} &= \frac{1}{\Sigma}\left[Q_1- \frac{\bt^2\,a^2 }{16}\,\frac{a^2 + (3a^2 + 2r^2)\cos2\theta}{(a^2 + r^2)^2}\right]\,,\\
		f^{(1,0,1)}_1&= -\frac{a^2\left(r^2 + (a^2-2r^2)\sin^2\theta\right)}{8 (a^2 + r^2)^2}\,,\\
		f_2^{(1,0,1)} &= \frac{a^2\left(r^2 - a^2\, \sin^2\theta\right)}{8(a^2 + r^2)^2}\,\\
		\mu^{(1,0,1)} &= - \frac{a^2\left(r^2 - (a^2 + 2r^2)\sin^2\theta\right)}{16\,\Sigma\,(a^2 + r^2)^2}\,,\\
		\nu^{(1,0,1)} &= - \frac{a^2\left(r^2 + a^2\,\sin^2\theta\right)}{16\,\Sigma\,(a^2 + r^2)^2}\,,\\
		\gamma_1^{(1,0,n)} &=\frac{\cos^2\theta}{\Sigma}\left[-  Q_{1}\,(r^2 + a^2) + \frac{\bt^2}{4}\frac{a^2\,r^2(2a^2 + r^2)\,\sin^2\theta}{(a^2 + r^2)^2}\right]d\phi \wedge d\psi \,. 
	\end{align}
\end{subequations}
At this point, we can combine the solutions into relevant forms
\begin{subequations}
	\label{eq:CForms101}
	\begin{align}
		\at_2 &= -\frac{a^2\,\bt^2}{4\,\sqrt2\,R_y}\left(\frac{(a^2-r^2)\,\sin^2\theta}{(a^2+r^2)^2}\, d\phi + \frac{r^2\,\cos^2\theta}{(a^2+r^2)^2}\,d\psi\right)\,,\\
		\omega &= \omega_0 + \frac{a^2\,\bt^2\,Q_5}{4\,\sqrt2\,R_y\,\Sigma}\,\left(\frac{\sin^2\theta}{a^2 + r^2}\,d\phi - \frac{r^2\,\cos^2\theta}{(a^2+ r^2)^2}\,d\psi\right)\,,
	\end{align}
\end{subequations}
and
\begin{align}
	\label{eq:CTheta2Form101}
	\ttt^2 &= -\frac{a^2\,\bt^2}{2\,\sqrt2\,R_y\,(a^2+ r^2)^2}\Big(r\,\sin^2\theta\, dr\wedge d\phi +\frac{r\,(a^2 -r^2)\,\cos^2\theta}{a^2 + r^2}\,dr\wedge d\psi\nonumber\\*
	&\quad  + (a^2 -r^2)\,\sin\theta\cos\theta\,d\theta\wedge d\phi - r^2\,\sin\theta \,\cos\theta\,d\theta\wedge d\psi \Big)
\end{align}
\paragraph{(1,1,1).}
We can similarly present the (1,1,1) solution
\begin{subequations}
	\label{eq:111}
	\begin{align}
		f^{(1,1,1)} &= -\frac{1}{4}\frac{1}{a^2 + r^2} \,,\\
		Z_1^{(1,1,1)} &= \frac{1}{\Sigma}\left[Q_1+ \frac{\bt^2\,a^2 }{16}\,\frac{a^2 - (3a^2 + 2r^2)\cos2\theta}{(a^2 + r^2)^2}\right]\,,\\
		f^{(1,1,1)}_1&= \frac{a^2\left(a^2 - (a^2+2r^2)\cos2\theta\right)}{8 (a^2 + r^2)^2}\,,\\
		f_2^{(1,1,1)} &= \frac{a^2\left(a^2 +r^2 - a^2\, \sin^2\theta\right)}{4(a^2 + r^2)^2}\,\\
		\mu^{(1,1,1)} &=  \frac{a^2 +r^2 -a^2\,\cos2\theta}{8\,\Sigma\,(a^2 + r^2)}\,,\\
		\nu^{(1,1,1)} &=  \frac{-a^2+\left(r^2 + a^2\right)\cos2\theta}{8\,\Sigma\,(a^2 + r^2)}\,,\\
		\gamma_1^{(1,1,n)} &=\frac{\cos^2\theta}{\Sigma}\left[-  Q_{1}\,(r^2 + a^2) - \frac{\bt^2}{4}\frac{a^2\,r^2(2a^2 + r^2)\,\sin^2\theta}{(a^2 + r^2)^2}\right]d\phi \wedge d\psi \,. 
	\end{align}
\end{subequations}
Recombining the solutions into forms gives
\begin{subequations}
	\label{eq:CForms111}
	\begin{align}
		\at_2 &= \frac{a^2\,\bt^2}{2\,\sqrt2\,R_y}\left(\frac{\sin^2\theta}{(a^2+r^2)}\, d\phi-  \frac{r^2\,\cos^2\theta}{(a^2+r^2)^2}\,d\psi\right)\,,\\
		\omega &= \omega_0 + \frac{\bt^2\,Q_5}{2\,\sqrt2\,R_y\,\Sigma}\,\left(\frac{(2a^2+ r^2)\sin^2\theta}{a^2 + r^2}\,d\phi + \frac{r^2\,\cos^2\theta}{(a^2+ r^2)^2}\,d\psi\right)\,,
	\end{align}
\end{subequations}
and
\begin{align}
	\ttt^2 &= -\frac{a^2\,\bt^2}{\sqrt2\,R_y\,(a^2+ r^2)^2}\Big(r\,\sin^2\theta\, dr\wedge d\phi - r \cos^2\,dr\wedge d\psi\nonumber\\*
	&\quad  - (a^2 +r^2)\,\sin\theta\cos\theta\,d\theta\wedge d\phi - r^2\,\sin\theta \,\cos\theta\,d\theta\wedge d\psi \Big)\,.
\end{align}

\paragraph{(2,0,1).} In the following examples we do not give the explicit values of $\gamma_1$, which can be determined by inverting the equations \eqref{eq:Cond2}.
\begin{subequations}
	\label{eq:201}
	\begin{align}
		f^{(2,0,1)} &= -\frac{2 a^4+a^2 \left(a^2+5 r^2\right) \,\sin ^2\theta +5
			a^2 r^2+3 r^4}{36 \left(a^2+r^2\right)^3}\,,\\
		Z_1^{(2,0,1)} &= \frac{1}{\Sigma}\left[Q_1+ \bt^2\frac{a^2 \left(a^2+r^2\right) \left(3 a^2+2 r^2\right)
			\cos (2 \theta )+a^4 \left(2 a^2+r^2\right) \sin ^2(2
			\theta )}{18 \left(a^2+r^2\right)^3 }\right]\,,\\
		f^{(2,0,1)}_1&=-\frac{a^2 \left(a^2 \left(\left(a^2-4 r^2\right) \cos (4
			\theta )+5 a^2-2 r^2\right)+\left(-6 a^4+10 a^2 r^2+4
			r^4\right) \cos (2 \theta )\right)}{72
			\left(a^2+r^2\right)^3}\,,\\
		f_2^{(2,0,1)} &= \frac{a^2 r^2 \left(a^2+r^2\right)+a^4 \sin ^2(\theta )
			\left(a^2 \cos (2 \theta )-2 a^2+3 r^2\right)}{18
			\left(a^2+r^2\right)^3}\,\\
		\mu^{(2,0,1)} &= \frac{a^4 \left(\left(a^2+5 r^2\right) \cos (4 \theta )+11
			a^2+7 r^2\right)-4 a^2 \left(a^2+r^2\right) \left(3
			a^2+2 r^2\right) \cos (2 \theta )}{288
			\left(a^2+r^2\right)^3 \,\Sigma}\,,\\
		\nu^{(2,0,1)} &=  \frac{1}{288
			\left(a^2+r^2\right)^3\,\Sigma}\Big[a^2 \big(12 a^2 \left(a^2+r^2\right) \cos
			(2 \theta )-11 a^4-25 a^2 r^2\nonumber\\*
			&\quad +\left(-a^4+5 a^2 r^2+2
			r^4\right) \cos (4 \theta )-10 r^4\big)\Big]\,.
	\end{align}
\end{subequations}
All fields in the supersymmetric ansatz can then be determined using these solutions. 

\paragraph{(2,1,1).}
\begin{subequations}
	\label{eq:211}
	\begin{align}
		f^{(2,1,1)} &= \frac{-8 a^4+a^2 \left(2 a^2+r^2\right) \cos (2 \theta
			)-15 a^2 r^2-6 r^4}{72 \left(a^2+r^2\right)^3}\,,\\
		Z_1^{(2,1,1)} &= \frac{1}{\Sigma} \left[Q_1+\bt^2\,\frac{a^4 \left(2 a^2+r^2\right) \cos ^2(2 \theta )}{18
			\left(a^2+r^2\right)^3 }\right]\,,\\
		f^{(2,1,1)}_1&=-\frac{a^2 \left(2 a^2+r^2\right) \cos (2 \theta )
			\left(a^2 (-\cos (2 \theta ))+a^2+2 r^2\right)}{36
			\left(a^2+r^2\right)^3}\,,\\
		f_2^{(2,1,1)} &= \frac{a^2 \left(a^4 \cos (4 \theta )+a^4-a^2 \left(2
			a^2+r^2\right) \cos (2 \theta )+5 a^2 r^2+2
			r^4\right)}{36 \left(a^2+r^2\right)^3}\,\\
		\mu^{(2,1,1)} &= \frac{3 \left(a^2+r^2\right)^3+a^2 \cos (2 \theta )
			\left(-5 a^4+a^2 \left(2 a^2+r^2\right) \cos (2 \theta
			)-10 a^2 r^2-4 r^4\right)}{72 \left(a^2+r^2\right)^3
			\left(a^2 \cos ^2(\theta )+r^2\right)}\,,\\
		\nu^{(2,1,1)} &=  \frac{1}{144\,\Sigma\, \left(a^2+r^2\right)^3}\Big[2 \left(5 a^6+10 a^4 r^2+9 a^2 r^4+3 r^6\right) \cos
			(2 \theta )\nonumber\\*
			& \quad -a^2 \left(8 a^4+\left(2 a^2+r^2\right)
			\left(a^2+2 r^2\right) \cos (4 \theta )+15 a^2 r^2+6
			r^4\right)\Big]\,.
	\end{align}
\end{subequations}

\paragraph{(2,2,1).}
\begin{subequations}
	\label{eq:221}
	\begin{align}
		f^{(2,2,1)} &= -\frac{a^2 \cos (2 \theta )+3 a^2+2 r^2}{8
			\left(a^2+r^2\right)^2}\,,\\
		Z_1^{(2,2,1)} &=\frac{1}{\Sigma} \left[Q_1+\bt^2\,\frac{a^4 \left(2 a^2+r^2\right) \sin ^2(2 \theta )-a^2
			\left(a^2+r^2\right) \left(3 a^2+2 r^2\right) \cos (2
			\theta )}{18 \left(a^2+r^2\right)^3}\right]\,,\\
		f^{(2,2,1)}_1&=-\frac{a^2 \left(-3 a^4+a^2 \left(a^2+2 r^2\right) \cos (4
			\theta )+2 \left(a^4+5 a^2 r^2+2 r^4\right) \cos (2
			\theta )\right)}{24 \left(a^2+r^2\right)^3}\,,\\
		f_2^{(2,2,1)} &= \frac{a^2 \left(a^2+r^2\right)^2+a^4 \cos ^2(\theta )
			\left(r^2-a^2 \cos (2 \theta )\right)}{6
			\left(a^2+r^2\right)^3}\,\\
		\mu^{(2,2,1)} &= \frac{-3 a^4 \cos (4 \theta )+15 a^4-4 a^2 \left(3 a^2+2
			r^2\right) \cos (2 \theta )+32 a^2 r^2+16 r^4}{96\,\Sigma\,
			\left(a^2+r^2\right)^2}\,,\\
		\nu^{(2,2,1)} &=  \frac{\left(3 a^2+2 r^2\right) \left(a^2 (\cos (4 \theta
			)-5)+4 \left(a^2+2 r^2\right) \cos (2 \theta
			)\right)}{96\,\Sigma\, \left(a^2+r^2\right)^2}\,.
	\end{align}
\end{subequations}

\paragraph{(2,1,2).}
\begin{subequations}
	\label{eq:212}
	\begin{align}
		f^{(2,1,2)} &= \frac{a^2 \left(a^4+6 a^2 r^2+2 r^4\right) \cos (2 \theta
			)-3 \left(a^2+2 r^2\right) \left(3 a^4+6 a^2 r^2+2
			r^4\right)}{384 \left(a^2+r^2\right)^4}\,,\\
		Z_1^{(2,1,2)} &=\frac{1}{\Sigma} \left[Q_1+\bt^2\,\frac{a^4 \left(a^4+6 a^2 r^2+2 r^4\right) \cos ^2(2
			\theta )}{96 \left(a^2+r^2\right)^4}\right]\,,\\
		f^{(2,1,2)}_1&=-\frac{a^2 \left(a^4+6 a^2 r^2+2 r^4\right) \cos (2 \theta
			) \left(a^2 (-\cos (2 \theta ))+a^2+2 r^2\right)}{192
			\left(a^2+r^2\right)^4}\,,\\
		f_2^{(2,1,2)} &=\frac{1}{384
			\left(a^2+r^2\right)^4}\Big[a^2 \big(a^6+6 a^4 r^2+28 a^2 r^4+a^4 \left(a^2+10
			r^2\right) \cos (4 \theta )\nonumber\\*
			&\quad -2 a^2 \left(a^4+6 a^2 r^2+2
			r^4\right) \cos (2 \theta )+8 r^6\big)\Big]\,\\
		\mu^{(2,1,2)} &= \frac{1}{768\,\Sigma\, \left(a^2+r^2\right)^4}\Big[-2
		\left(5 a^2+2 r^2\right) \left(a^3+2 a r^2\right)^2
		\cos (2 \theta )+8 r^8\nonumber\\*
		&\quad +9 a^8+38 a^6 r^2+50 a^4 r^4+32 a^2 r^6+a^4
			\left(a^4+6 a^2 r^2+2 r^4\right) \cos (4 \theta )\Big]\,,\\
		\nu^{(2,1,2)} &=\frac{1}{384\,\Sigma\,
			\left(a^2+r^2\right)^4 }\Big[ \big(2
			\left(5 a^8+22 a^6 r^2+26 a^4 r^4+16 a^2 r^6+4
			r^8\right) \cos (2 \theta )\nonumber\\*
			&\quad -a^2 \left(a^2+2 r^2\right)
			\left(9 a^4+18 a^2 r^2+\left(a^4+6 a^2 r^2+2 r^4\right)
			\cos (4 \theta )+6 r^4\right)\big)\Big]\,.
	\end{align}
\end{subequations}

\section{Conserved charges for $G_{-\frac12}^{+A}\kett{\dot A -}_k^{\rm NS}$ states}
\label{app:SecondState}

Let us analyse the state
\begin{align}
	&G_{-\frac12}^{+A}\kett{\dot A -}_k^{\rm NS} &&h^{\rm NS} = \frac{k}{2}+ \frac12\,, \quad j^{\rm NS} = -\frac{k}{2}+ \frac12\,,&& \hbn = - \jbn =  \frac{k }{2}-  \frac12\,,
\end{align}
in a bit more detail. 
We already mentioned that by shifting $k \to k+1$, the CFT charges of this state become those of the state \eqref{eq:StartState}, whose perturbation we constructed in Section~\ref{sec:Perturbation}.
By acting with the global symmetry generators we obtain the more general state
\begin{align}
	\label{eq:Dstate1}
	 \left(L_{-1}\right)^{n-1} \left(J_0^{+}\right)^m G_{-\frac12}^{+A}\kett{\dot A -}_k^{\rm NS}\,,
\end{align}
with
\begin{align}
	\label{eq:Dchargesstate1}
	h^{\rm NS} = \frac{k}{2}- \frac12+n\,, \qquad j^{\rm NS} = -\frac{k}{2}+ \frac12+m\,, \qquad \hbn = - \jbn =  \frac{k }{2}-  \frac12\,,
\end{align}
which is then tensored with the NS-NS vacuum state 
\begin{align}
	\label{eq:DFullStateNS}
	\left(\kett{--}_1^{\rm NS}\right)^{N_a}\, \left(\left(L_{-1}\right)^{n-1} \left(J_0^{+}\right)^m G_{-\frac12}^{+A}\kett{\dot A -}_k^{\rm NS} \right)^{N_b}\,, \qquad  N_b \ll N_a\,,
\end{align}
subject to the constraint 
\begin{align}
	\label{eq:DStrandBudget}
	N = N_a + k \, N_b\,.
\end{align}
The spectral flow to the Ramond sector gives a state with the charges
\begin{align}
	\label{eq:DRCharges}
	h^{\rm R} =  \frac{N}4 + N_b \left(m+n\right)\,, \qquad \hb^{\rm R} = \frac{N}{4}\,, \qquad j^{\rm R} = \frac{N_a+ N_b}{2} + N_b\, m\,, \qquad \jb^{\rm R} = \frac{N_a + N_b}{2}\,.
\end{align}
and non-zero momentum
\begin{align}
	\label{eq:DMomChargeCFT}
	n_P^{\rm R} =  h^{\rm R}- \hb^{\rm R} =  N_b \left(m+n\right)\,.
\end{align} 

On the supergravity side, the calculations proceed as in the main text: Except that one should shift $k \to k+1$. For example, the explicit solutions presented in Section~\ref{sec:ExpExam} correspond to a state with mode numbers $(2,0,n)$ and $(2,1,n)$.
The most important change appears in matching between the gravity and CFT moduli.
The regularity condition is 
\begin{align}
	\label{eq:DRegularity}
	Q_1\, Q_5 = a^2\, R_y^2 + x_{k-1,m,n}\, \frac{b^2}{2}\,.
\end{align} 
where we used \eqref{eq:btobt}.
The fact that $\bar J$ is always given by
\begin{align}
	\bar J = \frac{a^2\, R_y}{2}\,, 
\end{align}
 immediately suggest that the translation 
\begin{align}
	\label{eq:DTrans1}
	\frac{N_a+ N_b}{N} = \frac{ R_y^2\, a^2}{Q_1\,Q_5}\,,\qquad \frac{(k-1)\,N_b}{N} =x_{k-1,m,n}\,\frac{b^2\,R_y^2}{2\, Q_1\,Q_5}\,.
\end{align}
Using this result, one can check that the momentum obtained from the asymptotic expansion of $\cft$ indeed reproduces the CFT result \eqref{eq:DMomChargeCFT}. 
Similarly, in all explicit examples analysed, the gravitational expression for the angular momentum $J$ reproduces the value of $j$ in \eqref{eq:DRCharges}.

The identification \eqref{eq:DTrans1} suggests that one cannot take $a =0$ without setting both $N_a = N_b=0$. 
In fact, the limit  $N_a=0$ corresponds to
\begin{align}
	a^2 = \frac{x_{k-1,m,n}}{2\,(k-1)}\,b^2\,,
\end{align}
which would be the minimally allowed value for $a$ that has a well-defined holographic dual. 
However, from the gravity point of view, there is no objection to analysing values with lower values of $a$.
We hope to return to this issue in future work. 

\bibliographystyle{JHEP}

\bibliography{VS}

\end{document}